\documentclass[conference]{IEEEtran}
\IEEEoverridecommandlockouts
\usepackage{amssymb}
\usepackage{hyperref}
\usepackage{cite}
\usepackage{graphicx}
\usepackage{array}
\usepackage{multirow}
\usepackage{amsmath}
\usepackage{stfloats}
\usepackage{graphicx}
\usepackage{subfigure}
\usepackage{tabularx}
\usepackage{booktabs}
\usepackage{epsfig,epsf}
\usepackage{setspace}
\usepackage{bm}
\usepackage{textcomp}
\usepackage{algorithmic}
\usepackage{algorithm}
\usepackage{subfigure}
\usepackage{caption}
\usepackage{graphicx}
\usepackage{setspace}
\usepackage{url}
\usepackage{verbatim}
\usepackage{graphicx}
\usepackage{cite}
\usepackage{amssymb}
\usepackage{color}
\usepackage{xpatch}
\definecolor{shadecolor}{RGB}{0,0,255}
\definecolor{blue}{RGB}{0,0,255}

\newcommand{\ee}{\mathrm e}
\newcommand{\jj}{\mathrm j}

\DeclareMathOperator{\sinc}{sinc}

\newtheorem{corollary}{Corollary}
\newtheorem{remark}{Remark}
\usepackage{soul}
\usepackage{booktabs} 
\usepackage{tabularx}
\usepackage{array} 
\usepackage[table]{xcolor} 
\definecolor{tableblue}{RGB}{221,238,248} \newcolumntype{Y}{>{\raggedright\arraybackslash}X}

\makeatletter

\ExplSyntaxOn
\cs_new:Npn \bibColoredItems #1#2
{
	\clist_map_inline:nn {#2} { \cs_new:cpn {bib@colored@##1} {#1} } 
}
\ExplSyntaxOff

\newcommand\bib@setcolor[1]{%
	\ifcsname bib@colored@#1\endcsname
	\expanded{\noexpand\color{\csname bib@colored@#1\endcsname}}%
	\else
	\normalcolor
	\fi
}

\makeatother

\begin{document}
	
\title{Cell-Level Channel Shaping for Rydberg Atomic Quantum Receivers in Satellite Uplinks With Doppler-Enabled Superheterodyne Reception}

\author{Qihao Peng, 
         Qu Luo,~\IEEEmembership{Member,~IEEE},
         Kezhi Wang,~\IEEEmembership{Senior Member,~IEEE}, 
        Cunhua Pan,~\IEEEmembership{Senior Member,~IEEE}, \\
     
       Pei Xiao,~\IEEEmembership{Senior Member,~IEEE},
        Trung Q. Duong,~\IEEEmembership{Fellow,~IEEE},  Jiangzhou Wang,~\IEEEmembership{Fellow,~IEEE}.
   
		\thanks{Q. Peng, Q. Luo, and P. Xiao are affiliated with 5G and 6G Innovation Centre, Institute for Communication Systems (ICS) of the University of Surrey, Guildford, GU2 7XH, UK. (e-mail: \{q.peng, q.u.luo,p.xiao\}@surrey.ac.uk). } \\
           \thanks{C. Pan and J. Wang are with the National Mobile Communications Research Laboratory, Southeast University, Nanjing, China. (e-mail: cpan@seu.edu.cn;j.z.wang@seu.edu.cn).}  \\
           \thanks{K. Wang is with the Department of Computer Science, Brunel University of London, UB8 3PH London, U.K. (e-mail: Kezhi.Wang@brunel.ac.uk).}\\
        \thanks{T. Q. Duong is with the Faculty of Engineering and Applied Science, Memorial University, St. John’s, NL A1C 5S7, Canada and also with the School of Electronics, Electrical Engineering and Computer Science, Queen’s University Belfast, Belfast, U.K. (e-mail: tduong@mun.ca.)}\\

        }

	
	\maketitle

\begin{abstract}
In this paper, we propose a self-superheterodyne Rydberg uniform array receiver for satellite uplink communications, in which the Doppler shift naturally induced by satellite motion is exploited to generate the intermediate-frequency signal. We first develop a near-field local oscillator (LO) synthesis model and characterize the spatially varying LO electric field across the Rydberg vapor cells. Based on a vapor-cell-center approximation, a closed-form radio frequency (RF)-to-optical conversion is derived, establishing an explicit bridge between the incident satellite signal and the LO-induced cell-level response. The derived model reveals that the programmable LO serves as an analog-domain channel-shaping mechanism by controlling the cell-level transduction gain, phase response, and phase-matching behavior. Building upon this equivalent channel model, we formulate an LO design problem that maximizes the Shannon capacity of the effective channel, and develop an efficient optimization algorithm for the LO amplitudes and phases. Simulation results demonstrate that the vapor-cell transduction can reshape the effective channel, adjust the beam-pattern alignment, and moderately reduce the inter-user correlation under suitable LO configurations. Furthermore, the proposed LO design significantly improves the achievable capacity over benchmark schemes, offering a promising self-superheterodyne Rydberg architecture for future satellite communication systems.
\end{abstract}


\section{Introduction}
Sixth-generation (6G) networks are envisioned to provide ubiquitous, high-throughput connectivity that seamlessly integrates terrestrial and non-terrestrial 
networks into a unified architecture~\cite{dang2020should,wang2023road,jiang2021road}. However, it is challenging to realize this target by relying only on terrestrial infrastructure, as it remains economically or geographically infeasible across remote, rural, oceanic, and aerial regions. Satellite communications offer a promising solution to this challenge by exploiting space-based base stations (BSs), thereby achieving ubiquitous connectivity~\cite{KodheliCOMST,Saadnetwork}. Among these satellites, low Earth orbit (LEO) satellites have attracted extensive attention from academia and industry, since LEO satellites operate in lower orbits, resulting in less path loss and lower propagation delay \cite{su2019broadband,you2020massive}. Benefiting from these advantages, large-scale LEO constellations, e.g., Starlink~\cite{ChaudhryTVT} and OneWeb~\cite{kokkoniemi2024mission}, have 
been deployed to provide global broadband service, underscoring the importance of satellite communications in future wireless communication systems.

To simultaneously serve multiple users over the same time-frequency resources, multiple-input multiple-output (MIMO) techniques have been extensively investigated for LEO satellite systems~\cite{abdelsadek2022distributed,li2021downlink}, spanning waveform design~\cite{shen2022random,liu2024otfs,luo2026standardizingaffinefrequencydivision}, channel estimation~\cite{li2023channel,wang2022joint}, emerging antennas~\cite{zhu2024dynamic,khan2023ris}, and multiple access~\cite{11552766,huang2023deep,chu2020robust}. For waveform design, orthogonal time-frequency space (OTFS) modulation was proposed to exploit channel sparsity and improve robustness to high satellite mobility \cite{shen2022random,liu2024otfs}. Qu \emph{et al.} investigated affine frequency division multiplexing (AFDM) by employing chirp-based subcarriers 
with tailored preambles, thereby achieving full diversity in high-mobility scenarios~\cite{luo2026standardizingaffinefrequencydivision}. Furthermore, to acquire the channel state information under the long propagation delay and limited pilot budget, estimators have been designed for orthogonal frequency-division multiplexing (OFDM) \cite{li2023channel} and OTFS \cite{wang2022joint}, respectively. To further exploit the spatial degrees of freedom, reconfigurable apertures such as movable antenna arrays~\cite{zhu2024dynamic} and 
reconfigurable intelligent surfaces (RISs)~\cite{khan2023ris} were employed to reshape the propagation environment, improving system performance. Finally, multiple-access techniques were applied to improve high efficiency and fairness, including sparse code multiple access \cite{11552766}, rate-splitting multiple access (RSMA) \cite{huang2023deep}, and non-orthogonal multiple access (NOMA) \cite{chu2020robust}. 

Although these prior works have made remarkable contributions to satellite communications, their receivers still rely on conventional radio frequency (RF) technology and thus cannot overcome the most fundamental bottleneck of direct satellite access, namely the constrained link budget \cite{heo2023mimo}. In particular, the limited transmit power of handheld terminals, the substantial free-space path loss over the earth-satellite distance, and the sensitivity floor of conventional RF front-ends jointly render the received signal extremely weak, fundamentally limiting the achievable performance.

Quantum sensing offers a promising route to overcome the link-budget issue by exploiting quantum phenomena to detect weak RF signals~\cite{schlossberger2024rydberg,zhang2024rydberg,fancher2021rydberg}. In particular, Rydberg atoms possess large electric dipole moments and strong polarizability~\cite{bussey2022quantum}, enabling interaction with external fields over a broad spectral range from direct current (DC) to terahertz. Leveraging this sensitivity 
through electromagnetically induced transparency (EIT) and Autler-Townes splitting~\cite{GongWCM,peng2026rfparadigmrydbergatomic,11568941}, a weak incident RF signal is optically transduced and read out by a photodetector. Owing to these benefits, the resulting Rydberg atomic quantum receivers (RAQRs) have the potential to achieve sensitivity that bypasses the Chu limit governing conventional RF antennas, providing a promising solution for next-generation satellite networks \cite{11417150}.

Building on these appealing properties, RAQRs have drawn extensive research attention. To bridge the fields of communications and physics, Cui \emph{et al.} studied a magnitude-only model and proposed an iterative detection algorithm \cite{CuiJSACatomc}. To mitigate the error propagation caused by the iterative algorithm, phase alignment was proposed in \cite{peng2025risassistedatomicmimoreceiver}. To enhance the sensitivity of RAQRs, Gong \emph{et al.} derived an equivalent baseband model based on a superheterodyne architecture, demonstrating its superior performance to conventional RF receivers~\cite{Gongtcom}. This work was subsequently extended to MIMO architectures~\cite{gong2026rydbergatomicquantummimo}. Furthermore, a dynamic signal response model was 
developed in~\cite{11278503}. To address the transit-time-related bandwidth limitation, Chen \emph{et al.} investigated a wide-band RAQR by using a six-wave mixing approach \cite{chen2026widebandquantumtransductionrydberg}. Beyond communication, the sensing capabilities of RAQRs have been further demonstrated in angle-of-arrival estimation~\cite{11493506}. Furthermore, by leveraging the ultra-high sensitivity of RAQRs, Peng \emph{et al.} explored the potential of MIMO simultaneous wireless information and power transfer to enable the passive Internet of Things~\cite{11547166}.

Despite these advances, most existing studies focus on either a single vapor cell or a co-located vapor-cell array, where the local oscillator (LO) is modeled as a far-field plane wave with uniform amplitude and phase, an assumption that is often unrealistic in practice. Consequently, two important issues remain unresolved. First, the non-uniform amplitude and phase distribution of the LO field across a vapor-cell array, which becomes significant in the near-field regime, has not been rigorously characterized. Second, the LO has largely been treated as a passive transduction reference, while its potential as a programmable analog-domain resource for actively shaping the effective channel remains unexplored. These limitations are particularly pronounced in LEO satellite uplinks, where severe Doppler shifts, near-field LO propagation, and strong user spatial correlation jointly necessitate a more accurate transduction model and a controllable channel-shaping mechanism.

To fill these gaps, this paper develops a cell-level 
Rydberg atomic quantum (RAQ)-MIMO framework based on a near-field LO. We exploit the Doppler shift as a natural intermediate frequency (IF) to realize a self-superheterodyne RAQR architecture, derive a physically accurate cell-level RF-to-optical conversion matrix that captures the spatial non-uniformity of the field, and demonstrate that this matrix can be leveraged for channel shaping to enhance capacity. The main contributions are summarized as follows:
\begin{itemize}
\item \textbf{Doppler-enabled self-superheterodyne architecture.} 
We propose a self-superheterodyne Rydberg atomic uniform array receiver for satellite uplink communications. Unlike conventional superheterodyne architectures, which require an externally generated frequency offset to produce the IF signal, the proposed receiver exploits the Doppler shift inherently induced by satellite motion to generate the IF component. This establishes a physically grounded self-superheterodyne mechanism and provides a natural framework for satellite-borne Rydberg array reception.

\item \textbf{Cell-level RF-to-optical conversion matrix.} We develop a three-dimensional near-field model that jointly characterizes LO synthesis and RF-to-optical conversion across a Rydberg vapor-cell array. Unlike existing models, the proposed framework explicitly captures the near-field propagation from programmable LO elements to each vapor cell, as well as the resulting non-uniform amplitude and phase distributions within the cell. Based on this model, we derive the LO-induced amplitude, phase, and local phase-gradient terms, and obtain a rigorous cell-level conversion matrix that governs the transduction behavior of each vapor cell. Simulation results demonstrate that the proposed model captures field-amplitude and phase variations much more accurately than the conventional uniform-field approximation.

\item \textbf{Conversion-matrix-enabled channel shaping.} 
We establish a closed-form cell-level RF-to-optical conversion model under the vapor-cell-center approximation. The resulting model explicitly reveals how vapor-cell transduction maps incident satellite RF signals into an equivalent baseband MIMO channel. More importantly, it demonstrates that the programmable LO can act as an analog-domain channel-shaping mechanism by controlling the cell-dependent conversion gain, phase response, and phase-matching factor, thereby reconfiguring the effective channel prior to digital processing.

\item \textbf{Capacity maximization LO design.} Based on the derived equivalent channel model, we formulate a Shannon-capacity maximization problem for LO design and jointly optimize the LO amplitudes and phases under practical hardware constraints. Unlike conventional precoding schemes that operate on an unconstrained weighting matrix, the proposed approach directly configures the programmable LO through the equivalent Rydberg MIMO channel. Simulation results show that the vapor-cell response can reshape the receive beam pattern, improve alignment with the desired signal direction, and reduce inter-user correlation under appropriate LO configurations. Consequently, the proposed LO design achieves a higher Shannon capacity than benchmark schemes, demonstrating the effectiveness of cell-level channel shaping for Rydberg satellite receivers.
\end{itemize}

The remainder of this paper is organized as follows. Section~II presents the cell-level RAQ-MIMO system model and the self-superheterodyne architecture. Section~III derives the 
cell-level RF-to-optical conversion matrix and characterizes the channel-shaping mechanism. Section~IV formulates the 
capacity-maximization problem and develops the projected gradient ascent algorithm. Section~V presents simulation results, and Section~VI concludes the paper.

\textit{Notations:} Throughout this paper, italic letters, boldface lowercase letters, and boldface uppercase letters denote scalars, vectors, and matrices, respectively. The operators $(\cdot)^{\mathrm{T}}$, $(\cdot)^{\mathrm{H}}$, $(\cdot)^{*}$, and $(\cdot)^{-1}$ denote the transpose, Hermitian transpose, complex conjugate, and inverse, respectively. The determinant, trace, and expectation operators are denoted by $\det[\cdot]$, $\mathrm{tr}(\cdot)$, and $\mathbb{E}\{\cdot\}$, respectively. The notation $\mathrm{diag}\{\cdot\}$ forms a diagonal matrix, $\odot$ denotes the Hadamard product, and $[\mathbf{A}]_{r,k}$ denotes the $(r,k)$-th entry of $\mathbf{A}$. For a complex quantity, $\Re\{\cdot\}$, $\Im\{\cdot\}$, $|\cdot|$, and $\arg\{\cdot\}$ denote the real part, imaginary part, magnitude, and phase, respectively. The Euclidean norm is denoted by $\|\cdot\|$. The set of complex numbers is denoted by $\mathbb{C}$, and $\mathbf{I}_N$ denotes the $N\times N$ identity matrix. The notation $\mathcal{CN}(\boldsymbol{\mu},\mathbf{R})$ denotes a circularly symmetric complex Gaussian distribution with mean $\boldsymbol{\mu}$ and covariance matrix $\mathbf{R}$. Furthermore, $(\cdot)'$ denotes differentiation, $j=\sqrt{-1}$ is the imaginary unit, and $c$, $\epsilon_0$, $q$, and $a_0$ denote the speed of light, vacuum permittivity, elementary charge, and Bohr radius, respectively.

\section{System Model}
In this section, we develop a detailed near-field LO model and elucidate the relationship between the LO electric field and the output probe beam.
\begin{figure*}
    \centering
    \IfFileExists{UPAsat.png}{%
    \includegraphics[width=1\linewidth]{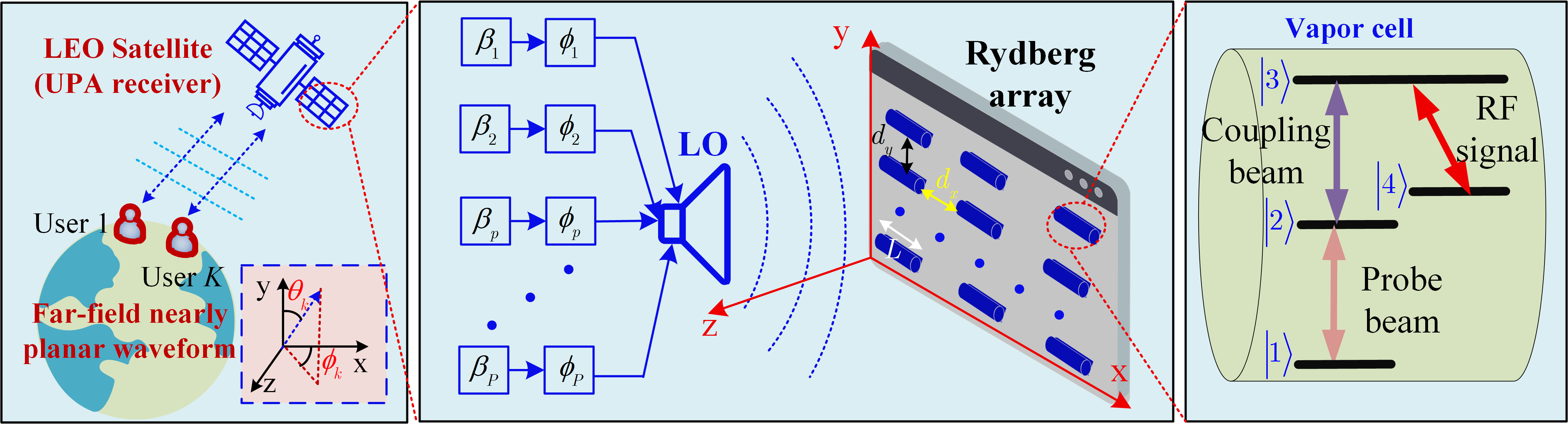}%
    }{%
    \fbox{\parbox[c][0.25\textheight][c]{1\linewidth}{\centering System model figure placeholder}}%
    }
    \caption{Rydberg atomic quantum satellite based on a four-level configuration.}
    \label{fig:placeholder}
\end{figure*}

\subsection{Two-Dimensional UPA Vapor-Cell Geometry}
As shown in Fig. \ref{fig:placeholder}, we consider a satellite-onboard two-dimensional UPA Rydberg atomic receiver. The array is located in the $x$--$y$ plane, while each vapor cell is placed along the $x$ direction. Let $l$ denote the local coordinate inside one vapor cell, where $0\le l\le L$ and $L$ is the physical length of a vapor cell. The coordinate of the point $l$ inside the $(m,n)$-th vapor cell is
\begin{equation}
\mathbf r_{m,n}(l)=
\begin{bmatrix}
(m-1)(L+d_x)+l\\
(n-1)d_y\\
0
\end{bmatrix},\quad 0\le l\le L.
\label{eq:cell_position}
\end{equation}
Here, $d_x$ denotes the physical gap between two neighboring vapor cells along the $x$ direction. Therefore, the actual $x$-direction array pitch is $D_x=L+d_x$.
\subsection{Electric Field Model}
\subsubsection{Electric Field of User}
The \(k\)-th user's signal is transmitted from Earth with a frequency of \(f_k = f_c + f_{D,k}\), where \(f_c\) is the carrier frequency and \(f_{D,k}\) is the Doppler shift between the user \(k\) and satellite. The arrival-direction unit vector of user $k$ is
\begin{equation}
\hat{\mathbf u}_k=u_k\hat{\mathbf x}+v_k\hat{\mathbf y}+w_k\hat{\mathbf z},
\label{eq:direction_vector}
\end{equation}
where
\begin{equation}
u_k=\sin\theta_k\cos\phi_k,\quad
v_k=\sin\theta_k\sin\phi_k,\quad
w_k=\cos\theta_k,
\label{eq:direction_cosines}
\end{equation}
and $u_k^2+v_k^2+w_k^2=1$.

Then, the electric field of user $k$ at the position $l$ inside the $(m,n)$-th vapor cell is
\begin{equation}
\label{eq:user_field}
\begin{split}
    \tilde E_{k,m,n}(l,t)&= \Re\Big\{
E_ke^{\jj\big[
2\pi f_kt + \theta_{m,n,k}(l) + \phi_k
\big]}\Big\},
\end{split}
\end{equation}
where $E_k$ denotes the received electric field near the receiver aperture and \(\theta_{m,n,k}(l) = k_cu_k[(m-1)D_x+l] + k_cv_k(n-1)d_y \) denotes the phase shift caused by the relative distance. $\phi_k$ is the common propagation and initial phase, and \( k_c =2\pi f_c/c  \approx 2\pi f_k/c  \) represents the wavenumber with the narrowband approximation.

\subsubsection{Electric Field of Programmable LO Phased Array}
We consider a programmable uniform linear array (ULA) LO phased array. The $p$-th LO element is located at 
\begin{equation} \mathbf r_{{\rm LO},p} = \begin{bmatrix} x_{{\rm LO}}\\ y_{{\rm LO}}+(p-1)d_{\rm LO}\\ z_{{\rm LO}} \end{bmatrix}, p\in \{1,\ldots,P\}. \label{eq:lo_ula_y_position} \end{equation} 
where $x_{{\rm LO}}$, \(y_{{\rm LO}}\), and \(z_{{\rm LO}}\) denote the reference coordinates of the LO array. $d_{\rm LO} = \frac{\lambda_\text{LO}}{2}$ is the inter-element spacing of the LO ULA with wavelength \(\lambda_\text{LO}\). The distance from the $p$-th LO element to the point $l$ inside the $(m,n)$-th vapor cell is \begin{equation} R_{p,m,n}(l) = \left\| \mathbf r_{{\rm LO},p} - \mathbf r_{m,n}(l) \right\|. \label{eq:lo_ula_distance} \end{equation} 
Assuming that the amplification coefficients and phases of LO are independently controllable, the complex amplifier coefficient of the $p$-th LO element is 
\begin{equation} 
a_{p} = \beta_{p}e^{\jj\phi_{p}}, \label{eq:lo_ula_excitation} 
\end{equation}
where $ \beta_{p}$ and $\phi_{p}$ denote the programmable amplification coefficient and phase shift of the $p$-th LO element, respectively. 

Using the spherical-wave near-field model, the electric field generated by the $p$-th LO element at the point $l$ inside the $(m,n)$-th vapor cell is given by 
\begin{equation}
\begin{split}
    \tilde E_{{\rm LO},p,m,n}(l,t) =  \Re\Big\{\frac{V_{\rm LO}a_{p}} {R_{p,m,n}(l)}e^{\jj \left[ 2\pi f_\text{LO}t - k_\text{LO}R_{p,m,n}(l)\right] }\},
\end{split}
\label{eq:lo_ula_element_field}
\end{equation} 
where \(f_\text{LO}\) and \(k_\text{LO}\) are the frequency and wavenumber, respectively. \(V_{\rm LO} = \sqrt{60P_{\rm LO}G_{\rm LO}}\) is the output voltage related to transmit power \(P_{\rm LO}\) and amplifier gain \(G_{\rm LO}\).
Therefore, the composite LO electric field at the point $l$ inside the $(m,n)$-th vapor cell is \begin{equation} 
\begin{split}
    \tilde E_{{\rm LO},m,n}(l,t) &= \sum_{p=1}^{P} \tilde E_{{\rm LO},p,m,n}(l,t) \triangleq \Re\Big\{\mathcal E_{m,n}(l) e^{\jj 2\pi f_\text{LO}t  }\Big\},
\end{split}
\label{eq:lo_ula_composite_field} 
\end{equation}
where \( \mathcal E_{m,n}(l)\) is the synthesized field envelope, given by
\begin{equation} 
\mathcal E_{m,n}(l) = V_{\rm LO} \sum_{p=1}^{P} \frac{a_{p}} {R_{p,m,n}(l)} e^{-\jj k_\text{LO}R_{p,m,n}(l) }.
\label{eq:lo_ula_composite_envelope} 
\end{equation} 
The received LO amplitude and phase of the \((m,n)\)-th vapor cell are defined as 
\begin{equation} 
E_{{\rm LO},m,n}(l) = \left| \mathcal  E_{m,n}(l) \right|, 
\label{eq:lo_ula_local_amplitude} 
\end{equation} 
and 
\begin{equation} 
\Phi_{{\rm LO},m,n}(l) = \arg \left\{ \mathcal  E_{m,n}(l) \right\}. 
\label{eq:lo_ula_local_phase}
\end{equation} 

\subsubsection{Electric Field of Superimposed Signal}
At the point $l$ inside the $(m,n)$-th vapor cell, the total RF electric field is
\begin{equation}
\tilde E_{{\rm RF},m,n}(l,t)=\Re\Big\{\tilde E_{{\rm LO},m,n}(l,t)+\sum_{k=1}^{K}\tilde E_{k,m,n}(l,t)\Big\}.
\label{eq:total_rf_field_passband}
\end{equation}
Defining \(f_\text{LO} = f_c\) and taking the LO phase as the reference, the amplitude of the superimposed RF signal is
\begin{equation}
\begin{split}
     E_{{\rm RF},m,n}(l,t) &= \Re\Big\{E_{{\rm LO},m,n}(l) + \\
    &\sum_{k=1}^{K}E_k \ee^{\jj [2\pi f_{D,k}t+\Delta\Phi_{k,m,n}(l)]}\Big\},
\end{split}
\label{eq:total_rf_field_lo_ref}
\end{equation}
where \(\Delta\Phi_{k,m,n}(l)\) denotes the user-LO relative phase, given by
\begin{equation}
\begin{aligned}
\Delta\Phi_{k,m,n}(l)
&=k_cu_k(m-1)D_x +k_cv_k(n-1)d_y\\
& +k_cu_kl+\phi_k-\Phi_{{\rm LO},m,n}(l).
\end{aligned}
\label{eq:relative_phase}
\end{equation}

Since the electric field from the user is much smaller than that of LO, we use the first-order approximation under the strong-LO condition. The RF-field envelope amplitude can be expressed as\footnote{Although the Doppler shift varies with time, user position, and satellite trajectory, it is treated as a fixed intermediate frequency within each channel block.}
\begin{equation}
\begin{split}
  E_{{\rm RF},m,n}(l,t)
&\approx
E_{{\rm LO},m,n}(l)\\
&+\sum_{k=1}^{K}E_k\cos[2\pi f_{D,k}t+\Delta\Phi_{k,m,n}(l)].
\end{split}
\label{eq:strong_lo_field_approx}
\end{equation}

\subsection{Four-Level Ladder System}
Here, we consider a four-level ladder system $|1\rangle\rightarrow |2\rangle\rightarrow |3\rangle\rightarrow |4\rangle$. The probe laser couples $|1\rangle\rightarrow |2\rangle$ in the \((m,n)\)-th vapor cell with Rabi frequency is
\begin{equation}
\Omega_{p,m,n}=\frac{\mu_{12}E_{p,m,n}}{\hbar},
\label{eq:omega_p}
\end{equation}
where \(\mu_{12}\) is the dipole moment. Similarly, the Rabi frequency for $|2\rangle\rightarrow |3\rangle$  and $|3\rangle\rightarrow |4\rangle$ are \(\Omega_{c,m,n} = \frac{\mu_{23}E_{c,m,n}}{\hbar}\) and \(\Omega_{\text{RF},m,n}(l,t) = \frac{\mu_{34}   E_{{\rm RF},m,n}(l,t)}{\hbar}\), where \(\mu_{23}\) and \(\mu_{34}\) are the corresponding dipole moments.

Defining $\boldsymbol{\rho}$ as the density matrix, the dynamics of the four-level transition scheme can be characterized by 
\begin{equation}
\begin{split}
    \frac{\text{d}\boldsymbol{\rho} }{\text{d} t} &= -j [\mathbf{H},\boldsymbol{\rho}] - \frac{1}{2}\{\boldsymbol{\Gamma},\boldsymbol{\rho} \} + \boldsymbol{\Lambda},\\
    \mathbf{H} &= \left[\begin{array}{cccc}
         0 & \frac{\Omega_p}{2} & 0& 0\\
          \frac{\Omega_p}{2}& \Delta_p & \frac{\Omega_c}{2} & 0\\
         0 &  \frac{\Omega_c}{2}& \Delta_p + \Delta_c &  \frac{\Omega_\text{RF}}{2} \\
         0& 0&\frac{\Omega_\text{RF}}{2} &\Delta_p + \Delta_c + \Delta_\text{RF}
    \end{array}\right], \\
    \boldsymbol{\Gamma} &= \text{diag}\{\gamma,\gamma+\gamma_2,\gamma+\gamma_3+\gamma_c,\gamma+\gamma_4\},\\
    \boldsymbol{\Lambda}& =  \text{diag}\{\gamma+\gamma_2\rho_{22}+\gamma_4\rho_{44},\gamma_{3}\rho_{33},0,0\},
\end{split}
\end{equation}
where \(\mathbf{H}\), \(\boldsymbol{\Gamma}\), and \(\boldsymbol{\Lambda}\) represent the Hamiltonian, the relaxation matrix, and the decay matrix, respectively. \(\rho_{mn}\) represents the \([m,n]\)-th element of \(\boldsymbol{\rho}\), \(\gamma_i\), \(i \in \{1,2,3,4\}\), is the spontaneous decay rate in the \(i\)-th level, \(\gamma\) and \(\gamma_c\) denote the relaxation rates associated with the atomic transition and collision, respectively.  $\Delta_p$, $\Delta_c$, and $\Delta_{\mathrm{RF}}$ are the corresponding detunings of the probe beam, coupling beam, and RF signal, respectively. For ease of derivations, we assume \(\gamma= \gamma_c = 0\) and decay rates of \(|3\rangle\) and \(|4\rangle\) are small enough to be ignored \cite{Gongtcom}. 

By setting \( \frac{\text{d}\boldsymbol{\rho} }{\text{d}t}  = 0\), we obtain the steady-state solution of density matrix \(\boldsymbol{\rho}\). Then, the measured susceptibility based on the probe beam is proportional to $\rho_{21}$, which is given by \cite{Gongtcom}
\begin{equation}\label{rho21}
    \begin{split}
 & \rho_{21}(\Omega_{\mathrm{RF}})=\Omega_{p}\times \\
 & \frac{A_{1}\Omega_{\mathrm{RF}}^{4}+A_{2}\Omega_{\mathrm{RF}}^{2}+A_{3}-j\left(B_{1}\Omega_{\mathrm{RF}}^{4}+B_{2}\Omega_{\mathrm{RF}}^{2}+B_{3}\right)}{C_{1}\Omega_{\mathrm{RF}}^{4}+C_{2}\Omega_{\mathrm{RF}}^{2}+C_{3}},
    \end{split}
\end{equation}
where \(A_1\), \(A_2\), \(A_3\), \(B_1\), \(B_2\), \(B_3\), \(C_1\), \(C_2\), \(C_3\) can be found in Appendix A of \cite{Gongtcom}. 

With the given RF signal, the power of the output probe beam is given by
\begin{equation}\label{outprobe}
\begin{split}
        P_{\text{out}}[\Omega_{\text{RF},m,n}(t)] &= P_{\text{in},m,n}  \exp\Bigg\{k_p D_{\Omega} \times \\
        & \int^L_0 \Im\Big[\rho_{21}(\Omega_{\text{RF},m,n}(l,t) )\Big]dl  \Bigg\},
\end{split}
\end{equation}
where \(P_\text{in}\) is the input power of the probe beam, \(D_{\Omega} = -\frac{2 N_0 \mu^2_{12}}{\epsilon_0\hbar \Omega_p}\) represents the constant term, and \(k_p = \frac{2\pi f_p}{c}\) means the wavenumber of the probe beam.

\section{Equivalent Channel Model}
In this section, we derive the closed-form expression of the output intensity of the probe beam and the equivalent channel model of UPA-based RAQR.
\subsection{Output Probe Intensity}
As seen from \eqref{outprobe}, it is challenging to obtain a closed-form expression for the output probe intensity due to the spatially varying RF Rabi frequency inside each vapor cell. To tackle this issue, we adopt a center-coordinate-based approximation. Specifically, since the physical length of each vapor cell is much smaller than the distance between the LO ULA and the Rydberg UPA, the LO amplitude variation inside the $(m,n)$-th vapor cell can be ignored. Therefore, the LO amplitude can be approximated by its value at the cell center $l=L/2$, i.e.,
\begin{equation}
E_{{\rm LO},m,n}(l)
\approx
E_{{\rm LO},m,n}^{c}
\triangleq
E_{{\rm LO},m,n}\left(\frac{L}{2}\right) = \left|
\mathcal E_{m,n}^{c}
\right|,
\label{eq:lo_amp_center_approx}
\end{equation}
where \(\mathcal E_{m,n}^{c}\) can be expressed as
\begin{equation}
\mathcal E_{m,n}^{c}
=
\mathcal E_{m,n}\left(\frac{L}{2}\right)
=
V_{\rm LO}
\sum_{p=1}^{P}
\frac{\beta_p}
{R_{p,m,n}^{c}}
e^{
\jj
\left(
\phi_p-k_{\rm LO}R_{p,m,n}^{c}
\right)}.
\label{eq:lo_center_envelope}
\end{equation}
\(R_{p,m,n}^{c}\) is given by
\begin{equation}
R_{p,m,n}^{c}
=
\sqrt{
(X_m^c)^2+
Y_{p,n}^2+
z_{\rm LO}^2
},
\label{eq:R_center_def}
\end{equation}
where \(X_m^c\) and \(Y_{p,n}\) are written as
\begin{equation}
X_m^c=x_{\rm LO}-(m-1)D_x-\frac{L}{2} \triangleq x_{\rm LO}- x^c_m ,
\label{eq:X_center_def}
\end{equation}
and
\begin{equation}
\begin{split}
    Y_{p,n}&=y_{\rm LO}+(p-1)d_{\rm LO}-(n-1)d_y\\ &\triangleq y_{\rm LO}+(p-1)d_{\rm LO} -y_n.
\end{split}
\label{eq:Y_center_def}
\end{equation}

Although the amplitude variation can be neglected, the LO phase variation inside the vapor cell should be retained, as the vapor-cell length may not be negligible compared with the RF wavelength. Therefore, we approximate the synthesized LO phase by the first-order Taylor expansion around the cell center as
\begin{equation}
\Phi_{{\rm LO},m,n}(l)
\approx
\Phi_{{\rm LO},m,n}^{c}
+
\zeta_{{\rm LO},m,n}^{c}
\left(
l-\frac{L}{2}
\right),
\label{eq:lo_phase_taylor_closed}
\end{equation}
where \(\zeta_{{\rm LO},m,n}^{c}\) means local LO phase slope, i.e.,
\begin{equation}
\begin{split}
    \zeta_{{\rm LO},m,n}^{c}
&=
\left.
\frac{d\Phi_{{\rm LO},m,n}(l)}
{dl}
\right|_{l=L/2}
=
\Im
\left\{
\frac{
\mathcal E_{m,n}'(L/2)
}
{
\mathcal E_{m,n}(L/2)
}
\right\}\\
&= \frac{
\mathcal E_{{\rm R},m,n}^{c}
\mathcal E_{{\rm I},m,n}^{\prime c}
-
\mathcal E_{{\rm I},m,n}^{c}
\mathcal E_{{\rm R},m,n}^{\prime c}
}
{
\left(\mathcal E_{{\rm R},m,n}^{c}\right)^2
+
\left(\mathcal E_{{\rm I},m,n}^{c}\right)^2
}.
\end{split}
\label{eq:lo_phase_slope_def}
\end{equation}
\(\mathcal E_{{\rm R},m,n}^{c}\) and  \(\mathcal E_{{\rm I},m,n}^{c}\) are the real and imaginary parts of $\mathcal E_{m,n}^{c}$, written as
\begin{equation}
\mathcal E_{{\rm R},m,n}^{c}
=
V_{\rm LO}
\sum_{p=1}^{P}
\frac{\beta_p}
{R_{p,m,n}^{c}}
\cos\chi_{p,m,n},
\label{eq:lo_real_center}
\end{equation}
and
\begin{equation}
\mathcal E_{{\rm I},m,n}^{c}
=
V_{\rm LO}
\sum_{p=1}^{P}
\frac{\beta_p}
{R_{p,m,n}^{c}}
\sin\chi_{p,m,n}.
\label{eq:lo_imag_center}
\end{equation}
\(\mathcal E_{{\rm R},m,n}^{\prime c}\) and \(\mathcal E_{{\rm I},m,n}^{\prime c}\) are
\begin{equation}
\begin{split}
    \mathcal E_{{\rm R},m,n}^{\prime c}
&=
V_{\rm LO}
\sum_{p=1}^{P}
\beta_p
\Big[
\frac{X_m^c}
{\left(R_{p,m,n}^{c}\right)^3}
\cos\chi_{p,m,n}\\
& -
\frac{k_{\rm LO}X_m^c}
{\left(R_{p,m,n}^{c}\right)^2}
\sin\chi_{p,m,n}
\Big],
\end{split}
\label{eq:lo_derivative_real_closed}
\end{equation}
and
\begin{equation}
\begin{split}
    \mathcal E_{{\rm I},m,n}^{\prime c}
&=
V_{\rm LO}
\sum_{p=1}^{P}
\beta_p
\Big[
\frac{X_m^c}
{\left(R_{p,m,n}^{c}\right)^3}
\sin\chi_{p,m,n}\\
&+
\frac{k_{\rm LO}X_m^c}
{\left(R_{p,m,n}^{c}\right)^2}
\cos\chi_{p,m,n}
\Big],
\end{split}
\label{eq:lo_derivative_imag_closed}
\end{equation}
where \(\chi_{p,m,n}\) is
\begin{equation}
\chi_{p,m,n}
=
\phi_p-k_{\rm LO}R_{p,m,n}^{c}.
\label{eq:chi_def}
\end{equation}
\(\Phi_{{\rm LO},m,n}^{c}\) represents the center LO phase, given by
\begin{equation}
\Phi_{{\rm LO},m,n}^{c}
=
\operatorname{atan2}
\left(
\mathcal E_{{\rm I},m,n}^{c},
\mathcal E_{{\rm R},m,n}^{c}
\right).
\label{eq:lo_center_phase_closed}
\end{equation}

By combining the center-amplitude approximation with the first-order phase expansion, the LO-referenced total RF field in \eqref{eq:total_rf_field_lo_ref} can be rewritten as
\begin{equation}
\begin{split}
     E_{{\rm RF},m,n}(l,t)
&\approx
E_{{\rm LO},m,n}^{c}
+ \Re\Big\{
\sum_{k=1}^{K}\Big[
E_ke^{\jj 2\pi f_{D,k}t}e^{\jj \Phi_{k,m,n}^{c}} \\
&\times e^{\jj \kappa_{k,m,n}^{c}
\left(
l-\frac{L}{2}
\right)}\Big]\Big\},
\end{split}
\label{eq:lo_ref_field_closed}
\end{equation}
where \(\Phi_{k,m,n}^{c}\) and \(\kappa_{k,m,n}^{c}\) are given by
\begin{equation}
\Phi_{k,m,n}^{c}
=
k_cu_kx_m^c
+
k_cv_ky_n
+
\phi_k
-
\Phi_{{\rm LO},m,n}^{c}
\label{eq:center_relative_phase_closed}
\end{equation}
and
\begin{equation}
\kappa_{k,m,n}^{c}
=
k_cu_k
-
\zeta_{{\rm LO},m,n}^{c}.
\label{eq:phase_slope_mismatch_closed}
\end{equation}
Therefore, the RF Rabi frequency becomes
\begin{equation}
\begin{split}
    \Omega_{{\rm RF},m,n}(l,t)
&\approx
\Omega_{{\rm LO},m,n}^{c}
+
\sum_{k=1}^{K}
\Omega_k\cos
\Big[
2\pi f_{D,k}t\\
&+
\Phi_{k,m,n}^{c}
+
\kappa_{k,m,n}^{c}
\left(
l-\frac{L}{2}
\right)
\Big],
\end{split}
\label{eq:rabi_closed}
\end{equation}
where \(\Omega_{{\rm LO},m,n}^{c}
=
\frac{\mu_{34}}
{\hbar}
E_{{\rm LO},m,n}^{c}\) and \(\Omega_k
=
\frac{\mu_{34}}
{\hbar}
E_k\).

Next, we linearize the imaginary part of $\rho_{21}(\Omega_{\rm RF})$ around the LO-biased operating point $\Omega_{{\rm LO},m,n}^{c}$, which is given by
\begin{equation}
\begin{split}
    \Im\left[
\rho_{21}
\left(
\Omega_{{\rm RF},m,n}(l,t)
\right)
\right]
&\approx
\Im\left[
\rho_{21}
\left(
\Omega_{{\rm LO},m,n}^{c}
\right)
\right]
\\
&+
g_{m,n}^{c}
\Delta\Omega_{m,n}(l,t),
\end{split}
\label{eq:rho_im_linearization}
\end{equation}
where \(\Delta\Omega_{m,n}(l,t)\) is written as
\begin{equation}
\begin{split}
    \Delta\Omega_{m,n}(l,t)
&=
\sum_{k=1}^{K}
\Omega_k
\cos
\Big[
2\pi f_{D,k}t
\\
&+\Phi_{k,m,n}^{c}
 + \kappa_{k,m,n}^{c}
\left(
l-\frac{L}{2}
\right)
\Big].
\end{split}
\label{eq:delta_rabi_closed}
\end{equation}
\(g_{m,n}^{c}\) is given by \eqref{gmnc}, at the bottom of this page.
\begin{figure*}[hb]
          \hrulefill
    \centering
        \begin{equation}\label{gmnc}
            \begin{split}
       g_{m,n}^{c}
=
-2\Omega_p\Omega_{{\rm LO},m,n}^{c}
\frac{
(B_1C_2-B_2C_1)(\Omega_{{\rm LO},m,n}^{c})^4
+
2(B_1C_3-B_3C_1)(\Omega_{{\rm LO},m,n}^{c})^2+
(B_2C_3-B_3C_2)
}
{
\left[
C_1(\Omega_{{\rm LO},m,n}^{c})^4
+
C_2(\Omega_{{\rm LO},m,n}^{c})^2
+
C_3
\right]^2}
    \end{split}
    \end{equation}
\end{figure*}

Substituting \eqref{eq:rho_im_linearization} into \eqref{outprobe}, we obtain
\begin{equation}
\begin{split}
    P_{{\rm out},m,n}(t)
\approx
P_{{\rm in},m,n}
e^{k_pD_{\Omega} L
\Im
\left[
\rho_{21}
\left(
\Omega_{{\rm LO},m,n}^{c}
\right)
\right]}\\
\times e^{k_pD_{\Omega}g_{m,n}^{c}
\int_0^L
\Delta\Omega_{m,n}(l,t)dl}.
\end{split}
\label{eq:pout_two_exp}
\end{equation}
Define the LO-biased direct current (DC) output probe power as
\begin{equation}
P_{0,m,n}^{c}
=
P_{{\rm in},m,n}
\exp
\left\{
k_pD_{\Omega} L
\Im
\left[
\rho_{21}
\left(
\Omega_{{\rm LO},m,n}^{c}
\right)
\right]
\right\}.
\label{eq:p0_def_closed}
\end{equation}
The closed-form expression of the remaining integral is
\begin{equation}
\begin{aligned}
&\int_0^L
\cos
\left[
2\pi f_{D,k}t+\Phi_{k,m,n}^{c}
+
\kappa_{k,m,n}^{c}
\left(
l-\frac{L}{2}
\right)
\right]dl
\\
&\quad
=
L
\sinc
\left(
\frac{\kappa_{k,m,n}^{c}L}{2}
\right)
\cos
\left(
2\pi f_{D,k}t+\Phi_{k,m,n}^{c}
\right),
\end{aligned}
\label{eq:cos_integral_closed}
\end{equation}
where $\sinc(x)=\sin(x)/x$. Therefore, the output probe power admits the closed-form approximation
\begin{equation}
\begin{aligned}
P_{{\rm out},m,n}(t)
&\approx
P_{0,m,n}^{c}
\exp
\Bigg\{
k_pD_{\Omega}g_{m,n}^{c}L
\sum_{k=1}^{K}
\Omega_k
\\
&\times\sinc
\left(
\frac{\kappa_{k,m,n}^{c}L}{2}
\right)
\cos
\left[
2\pi f_{D,k}t+
\Phi_{k,m,n}^{c}
\right]
\Bigg\}.
\end{aligned}
\label{eq:pout_closed_exp}
\end{equation}
For weak user signals, the exponential term in \eqref{eq:pout_closed_exp} can be further linearized, i.e., \(e^{x} \approx 1 + x \), yielding
\begin{equation}
P_{{\rm out},m,n}(t)
\approx
P_{0,m,n}^{c}
+
\Delta P_{{\rm out},m,n}(t),
\label{eq:pout_dc_ac}
\end{equation}
where the alternating current (AC) component is
\begin{equation}
\begin{aligned}
\Delta P_{{\rm out},m,n}(t)
&\approx
P_{0,m,n}^{c}
k_pD_{\Omega}g_{m,n}^{c}L
\sum_{k=1}^{K}
\Omega_k
\sinc
\left(
\frac{\kappa_{k,m,n}^{c}L}{2}
\right)
\\
&\quad\times
\cos
\left[
2\pi f_{D,k}t+
\Phi_{k,m,n}^{c}
\right].
\end{aligned}
\label{eq:pout_ac_closed}
\end{equation}

\subsection{Equivalent Channel}

Based on the closed-form AC probe power in \eqref{eq:pout_ac_closed}, we now derive the equivalent voltage output of each vapor cell and the corresponding MIMO signal model. After photodetection and DC-blocking, the DC optical component is removed and the AC photocurrent is directly obtained as
\begin{equation}
i_{{\rm AC},m,n}(t)
=
\mathcal R_{\rm pd}
\Delta P_{{\rm out},m,n}(t)
+
n_{i,m,n}(t),
\label{eq:ac_photocurrent}
\end{equation}
where $\mathcal R_{\rm pd}$ is the photodetector responsivity and $n_{i,m,n}(t)$ denotes the equivalent photocurrent noise after DC-blocking.

The AC photocurrent is then amplified by an electrical amplifier with gain $G_{\rm amp}$ and converted into voltage through a load resistance $R_0$. Hence, the output voltage is
\begin{equation}
v_{m,n}(t)
=
R_0G_{\rm amp}i_{{\rm AC},m,n}(t).
\label{eq:output_voltage_from_current}
\end{equation}
In this work, we assume a unit load resistance, i.e., $R_0=1~\Omega$. Therefore, we obtain
\begin{equation}
\begin{aligned}
v_{m,n}(t)
&\approx
R_0G_{\rm amp}\mathcal R_{\rm pd}
P_{0,m,n}^{c}
k_pD_{\Omega}g_{m,n}^{c}L
\sum_{k=1}^{K}
\Omega_k
\\
&\times
\sinc
\left(
\frac{\kappa_{k,m,n}^{c}L}{2}
\right)
\cos
\left[
2\pi f_{D,k}t+
\Phi_{k,m,n}^{c}
\right]
\\
&+
R_0G_{\rm amp}n_{i,m,n}(t).
\end{aligned}
\label{eq:real_voltage_output_simplified}
\end{equation}

To establish the communication-theoretic equivalent channel, we distinguish the wireless electric-field channel from the LO-shaped Rydberg transduction channel. The complex electric-field envelope impinging on the $(m,n)$-th vapor cell from user $k$ can be modeled as
\begin{equation}
 E_{k,m,n}(t)
=
\sqrt{P_{t,k}}
h_{m,n,k}
s_k(t),
\label{eq:wireless_field_channel}
\end{equation}
where $P_{t,k}$ is the transmit power of user $k$, $s_k(t)$ is the normalized transmitted symbol with $\mathbb E[|s_k(t)|^2]=1$, and $h_{m,n,k}$ is the actual wireless electric-field channel from user $k$ to the $(m,n)$-th vapor cell. Substituting \eqref{eq:wireless_field_channel} into \eqref{eq:real_voltage_output_simplified} and applying a mixing filter, we can obtain the real and imaginary parts of the received signal, which is given by
\begin{equation}
v_{m,n}(t)
\!=\!
\sum_{k=1}^{K}
w_{{\rm LO},m,n,k}^{\rm cell}
h_{m,n,k}
\sqrt{P_{t,k}}
s_k(t)
e^{\jj2\pi f_{D,k}t}
+
n_{m,n}(t),
\label{eq:cell_complex_voltage_output}
\end{equation}
where \(n_{m,n}(t) = R_0G_{\rm amp}n_{i,m,n}(t)\) is the equivalent noise and $w_{{\rm LO},m,n,k}^{\rm cell}$ denotes the LO-shaped Rydberg transduction coefficient, given by
\begin{equation}
\begin{aligned}
w_{{\rm LO},m,n,k}^{\rm cell}
&=
R_0G_{\rm amp}\mathcal R_{\rm pd}
P_{0,m,n}^{c}
k_pD_{\Omega}g_{m,n}^{c}L
\frac{\mu_{34}}{\hbar}
\\
&\quad\times
\sinc
\left(
\frac{\kappa_{k,m,n}^{c}L}{2}
\right)
\exp
\left\{
-\jj\Phi_{{\rm LO},m,n}^{c}
\right\}.
\end{aligned}
\label{eq:lo_cell_transduction}
\end{equation}
The coefficient $w_{{\rm LO},m,n,k}^{\rm cell}$ is determined by the LO ULA and the Rydberg-optical readout chain. Specifically, the LO ULA controls $E_{{\rm LO},m,n}^{c}$, $\Phi_{{\rm LO},m,n}^{c}$, and $\zeta_{{\rm LO},m,n}^{c}$, which further determine $P_{0,m,n}^{c}$, $g_{m,n}^{c}$, and the phase-matching factor $\sinc(\kappa_{k,m,n}^{c}L/2)$.

Let the two-dimensional vapor-cell index $(m,n)$ be mapped to the one-dimensional receive index
\begin{equation}
r=(n-1)M_x+m,
\qquad
r=1,\ldots,M_R,
\label{eq:cell_index_mapping}
\end{equation}
where $M_R=M_xM_y$ is the total number of vapor cells. Define the wireless electric-field channel matrix as
\begin{equation}
\mathbf H
\in
\mathbb C^{M_R\times K},
\qquad
[\mathbf H]_{r,k}
=
h_{m,n,k},
\label{eq:actual_wireless_channel_matrix}
\end{equation}
and the LO-shaped cell-level transduction matrix as
\begin{equation}
\mathbf W_{\rm LO}^{\rm cell}
\in
\mathbb C^{M_R\times K},
\qquad
[\mathbf W_{\rm LO}^{\rm cell}]_{r,k}
=
w_{{\rm LO},m,n,k}^{\rm cell}.
\label{eq:lo_cell_matrix}
\end{equation}
Then, by defining the Doppler-and-power diagonal matrix as \(\mathbf D(t)
=
\operatorname{diag}
\left\{
\sqrt{P_{t,1}}e^{\jj2\pi f_{D,1}t},
\ldots,
\sqrt{P_{t,K}}e^{\jj2\pi f_{D,K}t}
\right\}\), the cell-level equivalent output is given by
\begin{equation}
\tilde{\mathbf y}_{\rm cell}(t)
=
\left(
\mathbf W_{\rm LO}^{\rm cell}
\odot
\mathbf H
\right)
\mathbf D(t)
\mathbf s(t)
+
{\mathbf n}_{\rm cell}(t),
\label{eq:strict_cell_level_model}
\end{equation}
where $\odot$ denotes the Hadamard product, $\mathbf s(t)=[s_1(t),\ldots,s_K(t)]^{T}$, and \({\mathbf n}_{\rm cell}(t) \in \mathbb{C}^{M_R \times 1}\) collects the noise background, including quantum projection noise, photon shot noise \(2R_0^2G_{\mathrm{amp}}^2\mathcal{R}_{\mathrm{pd}}P^c_{\rm{out},r}qB\), and thermal noise \(4R_0G_{\mathrm{amp}}^2k_BTB\) \cite{peng2026signaldependentshotnoisemodeling}.

\begin{remark}[Separation between wireless propagation and LO-shaped transduction]
The equivalent cell-level model separates the actual wireless channel from the Rydberg-optical-electrical transduction induced by the programmable LO aperture. Specifically, $\mathbf H$ characterizes the physical propagation from the users to the vapor-cell array, and $\mathbf W_{\rm LO}^{\rm cell}$ characterizes the RF-to-optical transduction, including the LO-shaped Rydberg response, RF-to-optical conversion, and the intra-cell phase-matching factor. 
\end{remark}

\begin{remark}[LO-induced cell-level aperture shaping]
\label{rem:lo_aperture_shaping_corr}
As can be seen from \eqref{eq:strict_cell_level_model}, the programmable LO does not merely serve as a frequency reference. Instead, it acts as an analog-domain aperture-shaping mechanism by imposing a cell-dependent transduction coefficient on the received RF field. For the \(r\)-th vapor cell and user \(k\), the LO-shaped cell-level coefficient can be written in the generic form
\begin{equation}
w_{{\rm LO},r,k}^{\rm cell}
=
\Gamma_r^{\rm LO}
\xi_{r,k}
e^{-\jj\Phi_{{\rm LO},r}^{c}},
\label{eq:remark_lo_cell_weight}
\end{equation}
where
\begin{equation}
\xi_{r,k}
=
\sinc
\left(
\frac{
\left[
k_cu_k-\zeta_{{\rm LO},r}^{c}
\right]L
}{2}
\right),
\label{eq:remark_phase_matching_factor}
\end{equation}
and
\begin{equation}
\Gamma_r^{\rm LO}
=
R_0G_{\rm amp}\mathcal R_{\rm pd}
P_{0,r}^{c}
k_pD_{\Omega}g_{r}^{c}L
\frac{\mu_{34}}{\hbar}.
\label{eq:remark_gamma_lo}
\end{equation}

For a target direction \((u_0,v_0)\), an ideal LO-shaped receive beam should align the cell-level phases and simultaneously satisfy the intra-cell phase-matching condition. Specifically, the LO center phase should approximately satisfy
\begin{equation}
\Phi_{{\rm LO},r}^{c}
\approx
k_cu_0x_r^c+k_cv_0y_r+\phi_0.
\label{eq:remark_phase_design}
\end{equation}
In addition, the local LO phase slope should satisfy
\begin{equation}
\zeta_{{\rm LO},r}^{c}
\approx
k_cu_0.
\label{eq:remark_slope_design}
\end{equation}
Under \eqref{eq:remark_phase_design} and \eqref{eq:remark_slope_design}, the signal arriving from \((u_0,v_0)\) is phase-aligned across vapor cells and phase-matched within each vapor cell. Therefore, the programmable LO effectively reshapes the receive aperture in the analog domain.
\end{remark}

\begin{remark}[LO-induced channel correlation control]\label{LimitedCC}
The cell-level transductions affect the inter-user correlation of the equivalent channel. Let \(\mathbf h_k\in\mathbb C^{M_R\times1}\) denote the original channel vector of user \(k\). The effective channel after the vapor-cell transduction is
\begin{equation}
\mathbf h_{{\rm eff},k}
=
\mathbf w_k\odot\mathbf h_k,
\label{eq:remark_heff_user}
\end{equation}
where \(\mathbf w_k=[\mathbf W_{\rm LO}^{\rm cell}]_{:,k}\). The normalized correlation between users \(i\) and \(j\) becomes
\begin{equation}
\rho_{i,j}^{\rm eff}
=
\frac{
\left|
\sum_{r=1}^{M_R}
w_{r,i}^{*}w_{r,j}
h_{r,i}^{*}h_{r,j}
\right|
}{
\sqrt{
\sum_{r=1}^{M_R}|w_{r,i}|^2|h_{r,i}|^2
}
\sqrt{
\sum_{r=1}^{M_R}|w_{r,j}|^2|h_{r,j}|^2
}
}.
\label{eq:remark_eff_corr_expanded}
\end{equation}
Compared with the original-channel correlation, \eqref{eq:remark_eff_corr_expanded} shows that the vapor-cell response modifies each cell-level cross-user term through the additional product \(w_{r,i}^{*}w_{r,j}\). Therefore, a judicious LO design can improve the spatial separability of the effective channel. Furthermore, it is worth noting that the ability to reduce correlations is limited.
Specifically, substituting \eqref{eq:remark_lo_cell_weight} into the cross-user product gives
\begin{equation}
w_{r,i}^{*}w_{r,j}
=
|\Gamma_r^{\rm LO}|^2
\xi_{r,i}^{*}\xi_{r,j}.
\label{eq:remark_common_lo_cancel}
\end{equation}
This indicates that the synthesized LO phase map is cell-dependent but user-independent, and thus the common LO phase cannot directly provide the user-specific phase diversity needed to suppress inter-user correlation. Consequently, such a mechanism can reshape the effective aperture, but its spatial separability is limited in the absence of frequency-domain or time-domain separation. This suggests that the programmable LO should be regarded as an analog-domain aperture-shaping resource rather than an 
unconstrained user-specific precoder.
\end{remark}

\begin{corollary}[Far-field LO case]\label{farfiled}
When the LO source is located in the far field of the vapor-cell array, the LO field can be approximated as a plane wave. In this case, all vapor cells experience the same LO amplitude but different position-dependent phases, i.e., \( \mathcal E_{{\rm LO},r}^{c}
    =
    E_{\rm LO}
    e^{{\rm j}\Phi_{{\rm LO},r}^{c}}\),
where \(E_{\rm LO}\) is constant for all vapor cells. Furthermore, the LO phase variation inside each vapor cell can be ignored, and thus the cell-level conversion coefficient becomes \(w_{{\rm LO},r,k}^{\rm cell}
    =
    \Gamma_{\rm LO}
    e^{-{\rm j}\Phi_{{\rm LO},r}^{c}}\), \(\forall r,k\)
Therefore, the effective equivalent channel becomes
\begin{equation} \label{farHeff}
    \mathbf H_{\rm eff}
    =
    \Gamma_{\rm LO}\boldsymbol{\Phi}_{\rm LO}
    \mathbf H,
\end{equation}
\end{corollary}
where \(\boldsymbol{\Phi}_{\rm LO} =   {\rm diag}\{\left[ e^{-{\rm j}\Phi_{{\rm LO},1}^{c}}, \ldots, e^{-{\rm j}\Phi_{{\rm LO},M_R}^{c}} \right]\}\). This result confirms the simplified form presented in \cite{Gongtcom}. As shown in \eqref{farHeff}, a far-field LO introduces only a common phase rotation and a uniform conversion gain, rather than the ability for channel shaping.

\section{LO Design for Cell-Level Channel Shaping}
In this section, given the wireless electric-field channel, we aim to design the programmable LO coefficients for realizing the vapor cell-level channel shaping. 
\subsection{Problem Formulation}
Based on the equivalent channel in \eqref{eq:strict_cell_level_model}, Shannon capacity can be expressed as
\begin{equation}
\begin{split} 
C(\mathbf a) = \log_2\det \Big[ \mathbf I_K + \mathbf H_{\rm eff}^{\rm H}(\mathbf a) \mathbf R_n^{-1} \mathbf H_{\rm eff}(\mathbf a) \Big],
\end{split} 
\label{eq:capacity_objective} 
\end{equation}
where \(\mathbf{H}_{\rm eff}(\mathbf{a}) = \left(
\mathbf W_{\rm LO}^{\rm cell}
\odot
\mathbf H
\right)
\mathbf D\) is the effective channel with Doppler matrix \(\mathbf{D}\) and \(\mathbf R_n = \mathbb E \left\{ \mathbf n_{\rm cell} \mathbf n_{\rm cell}^{\rm H} \right\}\) denotes the covariance matrix of noise. \(\mathbf{a} = [\beta_{1}e^{\jj\phi_{1}},\ldots,\beta_{P}e^{\jj\phi_{P}}]^T\) collects the coefficients of LO. Therefore,  the capacity-maximizing LO design is formulated as 
\begin{subequations}
\label{optimizatin}
\begin{align}
\mathcal P_1:\quad \max_{\mathbf{a}} \quad \notag & C(\mathbf a) \\ 
{\rm s.t.}\quad & 0\le\beta_p\le 1, \quad p=1,\ldots,P, \label{cona} \\ 
& 0\le\phi_p<2\pi, \quad p=1,\ldots,P. \label{conb}
\end{align} 
\end{subequations}
Constraints \eqref{cona} and \eqref{conb} indicate that the amplifier gain and phase shift are limited. As can be seen from \eqref{optimizatin}, it is challenging to solve owing to the non-convex and generally intractable formulation.

\subsection{Proposed Projected-Gradient Design} 
Here, we focus on solving LO design by assuming perfect knowledge of channel \(\mathbf{H}\). To address this issue, we define 
\begin{equation}
\begin{split}
\mathbf S(\mathbf a) &= \mathbf R_n + \mathbf H_{\rm eff}(\mathbf a) \mathbf H_{\rm eff}^{\rm H}(\mathbf a) \\
 &= \mathbf R_n + \left(
\mathbf W_{\rm LO}^{\rm cell}
\odot
\mathbf H
\right)\left(
\mathbf W_{\rm LO}^{\rm cell}
\odot
\mathbf H
\right)^{\rm H}.
\end{split}
\label{eq:capacity_S}
\end{equation} 
Note that \(\mathbf{D}\mathbf{D}^{\rm H}\) is assumed to be identity matrix.
Then,  using Sylvester's determinant identity, i.e., \(\det(\mathbf I+\mathbf A\mathbf B) = \det(\mathbf I+\mathbf B\mathbf A)\), the determinant in \eqref{eq:capacity_objective} can be rewritten as 
\begin{equation} 
\begin{split}C(\mathbf a)= & \log_2\det \left[ \mathbf I_K + \mathbf H_{\rm eff}^{\rm H}(\mathbf{a}) \mathbf R_n^{-1} \mathbf H_{\rm eff}(\mathbf{a}) \right] \\ 
 = &\log_2 \det \left[ \mathbf I_{M_R} + \mathbf R_n^{-1} \mathbf H_{\rm eff}(\mathbf{a}) \mathbf H_{\rm eff}^{\rm H} (\mathbf{a})\right] \\ 
 = & \log_2\det[\mathbf{S}(\mathbf{a})]-\log_2\det [\mathbf{R}_n]. 
\end{split}
\label{eq:capacity_det_equivalence} 
\end{equation}

To solve problem $\mathcal P_1$, we derive the gradient of the capacity with respect to each real-valued LO control variable \(x\in\{\beta_p,\phi_p\}\), \(p=1,\ldots,P\). Then, the capacity gradient is given by 
\begin{equation} 
\begin{split}
    \frac{\partial C}{\partial x}& = 2\Re \left\{ \operatorname{tr} \left[ \mathbf Z^{\rm H} \frac{\partial\left(
\mathbf W_{\rm LO}^{\rm cell}
\odot
\mathbf H
\right)}{\partial x} \right] \right\},
\end{split}
\label{eq:capacity_gradient} 
\end{equation}
where \(\mathbf Z = \frac{1}{\ln 2} \mathbf S^{-1} \mathbf H_{\rm eff}\). Therefore, evaluating the capacity gradient reduces to deriving $\frac{\partial\mathbf W_{\rm LO}^{\rm cell}}{\partial x}$.

Owing to \(\left[ \mathbf W_{\rm LO}^{\rm cell} \right]_{r,k} = w_{r,k}\), we have 
\begin{equation} 
\left[ \frac{\partial\mathbf W_{\rm LO}^{\rm cell}} {\partial x} \right]_{r,k} = \frac{\partial w_{{\rm LO},r,k}^{\rm cell}}{\partial x}. 
\label{eq:W_element_derivative} 
\end{equation} 
According to the derived vapor cell-level equivalent model in \eqref{eq:remark_lo_cell_weight} and applying the product and chain rules, we have
\begin{equation} 
\begin{split} 
\frac{\partial w_{{\rm LO},r,k}^{\rm cell}} {\partial x} & = e^{-\jj\Phi_{{\rm LO},r}^{c}} \Bigg[ \sinc \left( \frac{ \left[ k_cu_k-\zeta_{{\rm LO},r}^{c} \right]L }{2} \right) \frac{\partial\Gamma_r^{\rm LO}}{\partial x} \\ 
& - \Gamma_r^{\rm LO} \frac{L}{2} \sinc' \left( \frac{ \left[ k_cu_k-\zeta_{{\rm LO},r}^{c} \right]L }{2} \right) \frac{\partial\zeta_{{\rm LO},r}^{c}}{\partial x} \\ 
& - \jj\Gamma_r^{\rm LO} \sinc \left( \frac{ \left[ k_cu_k-\zeta_{{\rm LO},r}^{c} \right]L }{2} \right) \frac{\partial\Phi_{{\rm LO},r}^{c}}{\partial x} \Bigg]. \end{split} 
\label{eq:w_LO_rk_derivative} 
\end{equation}
The remaining derivatives in \eqref{eq:w_LO_rk_derivative} are given below.
\begin{equation} 
\begin{split} 
\frac{\partial\Gamma_r^{\rm LO}}{\partial x}& =  R_0G_{\rm amp}\mathcal R_{\rm pd} k_pD_{\Omega}L \frac{\mu_{34}}{\hbar} P_{0,r}^{c} \\ 
&\times \left[ k_pD_{\Omega}L \left(g_r^{c}\right)^2 + \left. \frac{ \partial^2\Im[\rho_{21}(\Omega)] }{ \partial\Omega^2 } \right|_{\Omega=\Omega_{{\rm LO},r}^{c}} \right] \frac{\partial\Omega_{{\rm LO},r}^{c}}{\partial x}, 
\end{split} 
\label{eq:Gamma_LO_derivative} 
\end{equation}
where
\begin{equation} 
\frac{\partial\Omega_{{\rm LO},r}^{c}}{\partial x} = \frac{\mu_{34}}{\hbar} \frac{ \Re \left\{ \left( \mathcal E_{{\rm LO},r}^{c} \right)^{*} \frac{\partial\mathcal E_{{\rm LO},r}^{c}} {\partial x} \right\} }{ \left| \mathcal E_{{\rm LO},r}^{c} \right| },
\label{eq:Omega_LO_derivative} 
\end{equation}
and 
\begin{equation}
\begin{split}
        \mathcal E_{{\rm LO},r}^{c} &= \mathcal E_{m,n}^{c},  \\
        \frac{\partial\mathcal E_{{\rm LO},r}^{c}} {\partial \beta_p} &= \frac{V_{\rm LO}}{R^c_{p,m,n}}e^{\jj \chi_{p,m,n}}, \\
         \frac{\partial\mathcal E_{{\rm LO},r}^{c}} {\partial \phi_p} &= \jj \frac{\beta_p V_{\rm LO}}{R^c_{p,m,n}}e^{\jj \chi_{p,m,n}}.
\end{split}
\end{equation}

\begin{equation} 
\begin{split} & \sinc' \left( \frac{ \left[ k_cu_k-\zeta_{{\rm LO},r}^{c} \right]L }{2} \right) \\ 
= &\frac{ \frac{ \left[ k_cu_k-\zeta_{{\rm LO},r}^{c} \right]L }{2} \cos \left( \frac{ \left[ k_cu_k-\zeta_{{\rm LO},r}^{c} \right]L }{2} \right) - \sin \left( \frac{ \left[ k_cu_k-\zeta_{{\rm LO},r}^{c} \right]L }{2} \right) }{ \left[ \frac{ \left[ k_cu_k-\zeta_{{\rm LO},r}^{c} \right]L }{2} \right]^2 }. 
\end{split} 
\label{eq:sinc_prime_explicit} 
\end{equation}

\begin{equation} 
\frac{\partial\zeta_{{\rm LO},r}^{c}}{\partial x} = \Im \left\{ \frac{ \mathcal E_{{\rm LO},r}^{c} \frac{ \partial \left(\mathcal E_{{\rm LO},r}^{c}\right)' }{ \partial x } - \left(\mathcal E_{{\rm LO},r}^{c}\right)' \frac{ \partial\mathcal E_{{\rm LO},r}^{c} }{ \partial x } }{ \left( \mathcal E_{{\rm LO},r}^{c} \right)^2 } \right\}. 
\label{eq:zeta_LO_derivative} 
\end{equation}

\begin{equation}
\frac{\partial\Phi_{{\rm LO},r}^{c}}{\partial x} = \Im \left\{ \frac{ 1 }{ \mathcal E_{{\rm LO},r}^{c} } \frac{ \partial\mathcal E_{{\rm LO},r}^{c} }{ \partial x } \right\}. \label{eq:Phi_LO_derivative} 
\end{equation}

To accelerate the convergence of the gradient ascent, a step size \(\eta^{(i)}\) is introduced at the \(i\)-th iteration. Meanwhile, since the capacity function is highly nonlinear with respect to the LO excitation vector, a fixed large step size may lead to an excessive update, causing the algorithm to skip over a local stationary point. To tackle this issue, we employ a backtracking line search to adaptively determine \(\eta^{(i)}\). Specifically, at each iteration, the step size is initialized as
\begin{equation}
\eta^{(i)}=\eta_0,
\label{eq:eta_initialization}
\end{equation}
where \(\eta_0>0\) is the initial trial step size. Given the capacity gradients \(\nabla_{\boldsymbol\beta}C^{(i)}\) and \(\nabla_{\boldsymbol\phi}C^{(i)}\), the tentative LO amplitude and phase updates are given by
\begin{equation}
\widetilde{\boldsymbol\beta}^{(i+1)}
=
\boldsymbol\beta^{(i)}
+
\eta^{(i)}
\nabla_{\boldsymbol\beta}C^{(i)},
\label{eq:beta_tentative_update}
\end{equation}
and
\begin{equation}
\boldsymbol\phi^{(i+1)}
=
{\rm mod}
\left(
\boldsymbol\phi^{(i)}
+
\eta^{(i)}
\nabla_{\boldsymbol\phi}C^{(i)},
2\pi
\right).
\label{eq:phi_tentative_update}
\end{equation}
Then, the tentative amplitude vector is projected onto the feasible set as
\begin{equation}
\boldsymbol\beta^{(i+1)}
=
\Pi_{\mathcal B}
\left(
\widetilde{\boldsymbol\beta}^{(i+1)}
\right),
\label{eq:beta_projection_update}
\end{equation}
where
\begin{equation}
\mathcal B
=
\left\{
\boldsymbol\beta:
0\le \beta_p\le 1
\right\}.
\label{eq:beta_feasible_set}
\end{equation}
The corresponding tentative LO excitation vector is then reconstructed as
\begin{equation}
\mathbf a^{(i+1)}
=
\boldsymbol\beta^{(i+1)}
\odot
e^{\jj\boldsymbol\phi^{(i+1)}}.
\label{eq:a_tentative_update}
\end{equation}

The gradient ascent-based LO design is detailed in Algorithm \eqref{alg:lo_capacity_gradient}. Specifically, we initialize the LO design with \(\mathbf{a}^{(i)}\) and calculate the cell-level LO coefficient \(\{w_{r,k}^{(i)}\}\), \(r \in \{1, M\}\), \(\forall k \in \{1,\ldots,K\}\), and corresponding capacity \(C^{(i)}\). In the \(i\)-th iteration, we obtain the gradient and update the   LO amplitudes and phases with step \(\eta^{(i)}\). Furthermore, the trial step size is accepted only when the updated LO vector yields a non-decreasing capacity, i.e., \(C^{(i+1)}\ge C^{(i)}\). Otherwise, the step size is reduced by using 
\begin{equation}
\eta^{(i)}
\leftarrow
\tau_{\eta}\eta^{(i)},
\qquad
0<\tau_{\eta}<1.
\label{eq:eta_backtracking_update}
\end{equation}
After that, the LO update, projection, and capacity evaluation are repeated. In this way, the algorithm preserves the accepted ascent while avoiding movements caused by an overly large step size.

\begin{algorithm}[t]
\caption{Gradient Ascent-based LO Design for Capacity Maximization}
\label{alg:lo_capacity_gradient}
\small
\begin{algorithmic}[1]
\STATE Initialize the LO coefficient vector
\(\mathbf a^{(1)}=\boldsymbol\beta^{(1)}\odot e^{\jj\boldsymbol\phi^{(1)}}\), the initial step size
\(\eta_1\), the step reduction factor \(0<\tau_{\eta}<1\), and the error tolerance \(\epsilon_{\rm stop}\). Set \(i=1\) and \(C^{(0)} = 0\).

\STATE  For the given LO vector \(\mathbf a^{(i)}\), calculate the cell-level LO coefficient \(\{w_{r,k}^{(i)}\}\), \(r \in \{1, \ldots,M_xM_y\}\), \(\forall k \in \{1,\ldots,K\}\), and construct
\(\mathbf W_{\rm LO}^{\rm cell,(i)}=[w_{r,k}^{(i)}]\).

\STATE Calculate the equivalent channel \(\mathbf H_{\rm eff}^{(i)}\) and the Shannon capacity \(C^{(i)}\).

\WHILE{\(\frac{|C^{(i+1)}-C^{(i)}|}
{\max\{C^{(i)},1\}}
\ge
\epsilon_{\rm stop}\)}

\STATE Calculate
\(\partial\mathbf W_{\rm LO}^{\rm cell,(i)}/\partial\beta_p\) and
\(\partial\mathbf W_{\rm LO}^{\rm cell,(i)}/\partial\phi_p\),
\(p=1,\ldots,P\), using
\eqref{eq:w_LO_rk_derivative}--\eqref{eq:Phi_LO_derivative}, obtain the capacity gradients \(\frac{\partial C^{(i)}}{\partial x}\), \(x \in \{\beta_p,\phi_p\}\), \(\forall p \in \{1,\ldots,P\}\).

\STATE Set \(\eta^{(i)}=\eta_1\), and update the LO's coefficients of amplitudes and phases by using \eqref{eq:beta_tentative_update}, \eqref{eq:phi_tentative_update}, and \eqref{eq:beta_projection_update}.
\STATE Construct the tentative LO vector
\(
\mathbf a^{(i+1)}
=
\boldsymbol\beta^{(i+1)}
\odot
e^{\jj\boldsymbol\phi^{(i+1)}}.
\) and compute  \(\mathbf W_{\rm LO}^{\rm cell,(i+1)}\),
\(\mathbf H_{\rm eff}^{(i+1)}\), and \(C^{(i+1)}\).

\WHILE{\(C^{(i+1)}<C^{(i)}\)}
\STATE Reduce the step size as
\(\eta^{(i)}\leftarrow\tau_{\eta}\eta^{(i)}\), and repeat Steps 5-7.
\ENDWHILE
\STATE Update \(i\leftarrow i+1\).
\ENDWHILE
\end{algorithmic}
\end{algorithm}

\section{Simulation Results}
In this section, we first evaluate the gap between our derived output voltage and that based on the Lindblad master equation. Then, based on our derived model, we investigate the benefits of cell-level channel shaping relying on our proposed algorithm. 

\begin{table*}[t]
\centering
\caption{Simulation Parameter Settings}
\label{tab:simulation_parameters}
\footnotesize
\renewcommand{\arraystretch}{1.12}
\setlength{\tabcolsep}{3pt}

\begin{minipage}[t]{0.3\textwidth}
\centering
\rowcolors{2}{gray!4}{white}
\begin{tabularx}{\linewidth}{@{}Y c c@{}}
\rowcolor{tableblue}
\multicolumn{3}{c}{\textbf{Satellite Parameters}}\\
\toprule
\textbf{Parameter} & \textbf{Value} & \textbf{Unit}\\
\midrule
Carrier frequency & \(f_c=6.9458\) & GHz\\
LO frequency & \(f_{\rm LO}=f_c\) & GHz\\
LEO altitude & \(h_{\rm sat}=550\) & km\\
Number of users & \(K=3\) & --\\
Transmit power & \(P_t=4\) & W\\
Receive gain & \(G_r= 0\) & dBi\\
Transmit gain & \(G_t=5\) & dBi\\
Doppler shifts & \(150\) & kHz\\
Rician factor & \(R_{{\rm R},k}=10\) & dB\\
\bottomrule
\end{tabularx}
\end{minipage}
\hfill
\begin{minipage}[t]{0.38\textwidth}
\centering
\rowcolors{2}{gray!4}{white}
\begin{tabularx}{\linewidth}{@{}Y c c@{}}
\rowcolor{tableblue}
\multicolumn{3}{c}{\textbf{Rydberg Cell and Optical Parameters}}\\
\toprule
\textbf{Parameter} & \textbf{Value} & \textbf{Unit}\\
\midrule
Probe wavelength & \(\lambda_p=852\) & nm\\
Coupling wavelength & \(\lambda_{\rm coup}=510\) & nm\\
Probe power & \(P_{\rm in}=20.7\) & \(\mu\)W\\
Coupling power & \(P_{\rm coup}=17\) & mW\\
Quantum efficiency & \(\mathcal R_{\rm pd}=0.8\) & --\\
\(|1\rangle\)--\(>|2\rangle\) Dipole moment  & \(\mu_{12}=2.2327qa_0\) & C/m\\
\(|2\rangle\)--\(>|3\rangle\) Dipole moment & \(\mu_{23}=0.0226qa_0\) & C/m\\
\(|3\rangle\)--\(>|4\rangle\) Dipole moment & \(\mu_{34}=1443.45qa_0\) & C/m\\
Atomic density & \(N_0=4.89\times10^{10}\) & cm\(^{-3}\)\\

\bottomrule
\end{tabularx}
\end{minipage}
\hfill
\begin{minipage}[t]{0.3\textwidth}
\centering
\rowcolors{2}{gray!4}{white}
\begin{tabularx}{\linewidth}{@{}Y c c@{}}
\rowcolor{tableblue}
\multicolumn{3}{c}{\textbf{Other Parameters}}\\
\toprule
\textbf{Parameter} & \textbf{Value} & \textbf{Unit}\\
\midrule
LO elements & \(P=16\) & --\\
LO spacing & \(d_{\rm LO}=\lambda_{\rm LO}/2\) & m\\
LO-array position & \(z_{\rm LO}=-2\) & m\\
LO gain & \(G_{\rm LO}=1\) & --\\
LO power & \(P_{\rm LO}= 10\) & dBm\\
Bandwidth & \(B=100\) & kHz\\
Load & \(R_0=1\) & \(\Omega\)\\
Voltage gain & \(G_{\rm amp}=30\) & dB\\
Noise temperature & \(T=100\) & K\\
\bottomrule
\end{tabularx}
\end{minipage}

\end{table*}

\subsection{Parameter Settings}
Consider a four-level excitation scheme 
\(6{\rm S}_{1/2}\rightarrow 6{\rm P}_{3/2}\rightarrow 47{\rm D}_{5/2}\rightarrow 48{\rm P}_{3/2}\). 
Unless otherwise stated, the detailed parameter settings are listed in 
Table~\ref{tab:simulation_parameters}. The receiver employs an 
\(M_x\times M_y = 4\times4\) array of vapor cells, each of length \(L = 4\) cm, 
with inter-cell spacings along the \(x\)- and \(y\)-axes set to 
\(d_x = d_y = \lambda_c/2\). The RF channel of user \(k\) is modeled as a Rician 
channel, given by
\[
\mathbf{h}_{{\rm RF},k} = \sqrt{\tfrac{R_{{\rm R},k} L_{{\rm loss},k}}{R_{{\rm R},k}+1}}\,
\bar{\mathbf{h}}_{{\rm RF},k} 
+\sqrt{\tfrac{L_{{\rm loss},k}}{R_{{\rm R},k}+1}}\,
\tilde{\mathbf{h}}_{{\rm RF},k} \in \mathbb{C}^{M_R \times 1},
\]
where \(R_{{\rm R},k}\) is the Rician factor and 
\(L_{{\rm loss},k} = P_{t}G_tG_r\left(\tfrac{\lambda_c}{4\pi d_k}\right)^2\) 
denotes the large-scale fading of user \(k\). \(d_k\) denotes the distance between the user \(k\) and the satellite . \(\bar{\mathbf{h}}_{{\rm RF},k}\) is the deterministic LoS steering vector determined by the angle of arrival of user \(k\), and \(\tilde{\mathbf{h}}_{{\rm RF},k}\sim\mathcal{CN}(\mathbf{0},\mathbf{I}_{M_R})\) is the NLoS component. Similarly, the electric-field channel is 
expressed as
\[
\mathbf{h}_k = E_k\left(\sqrt{\tfrac{R_{{\rm R},k}}{R_{{\rm R},k}+1}}\,
\bar{\mathbf{h}}_{{\rm RF},k} 
+\sqrt{\tfrac{1}{R_{{\rm R},k}+1}}\,
\tilde{\mathbf{h}}_{{\rm RF},k}\right) \in \mathbb{C}^{M_R \times 1},
\]
where \(E_k = \tfrac{\sqrt{60P_tG_t}}{d_k}\) denotes the electric-field amplitude.

\subsection{Derivation Validation}
\begin{figure*}[t]
	\centering
	\subfigure[Amplitude.]{
		\includegraphics[width=3.25in]{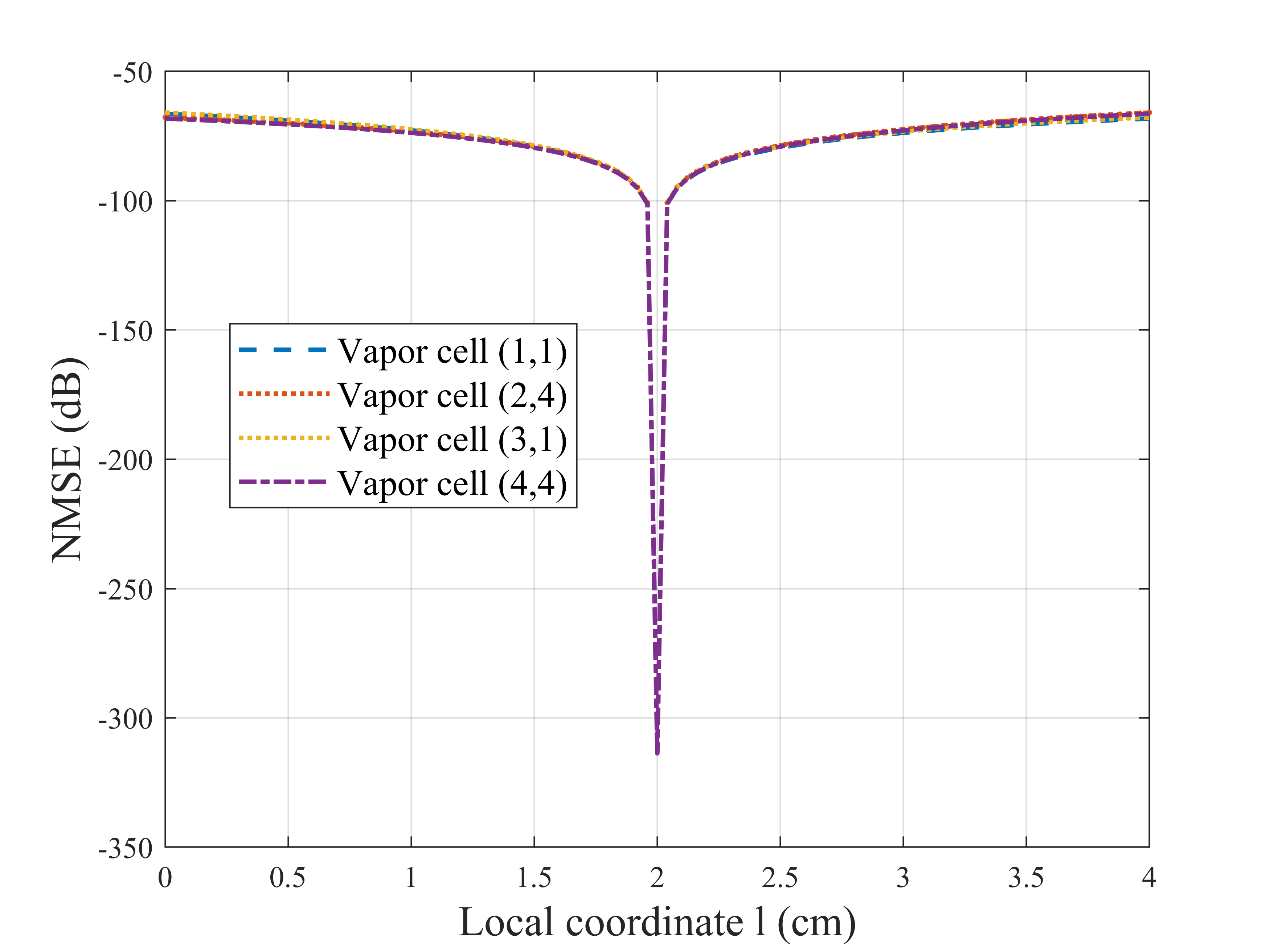}}\hspace{10mm}
	\subfigure[Phase.]{
		\includegraphics[width=3.25in]{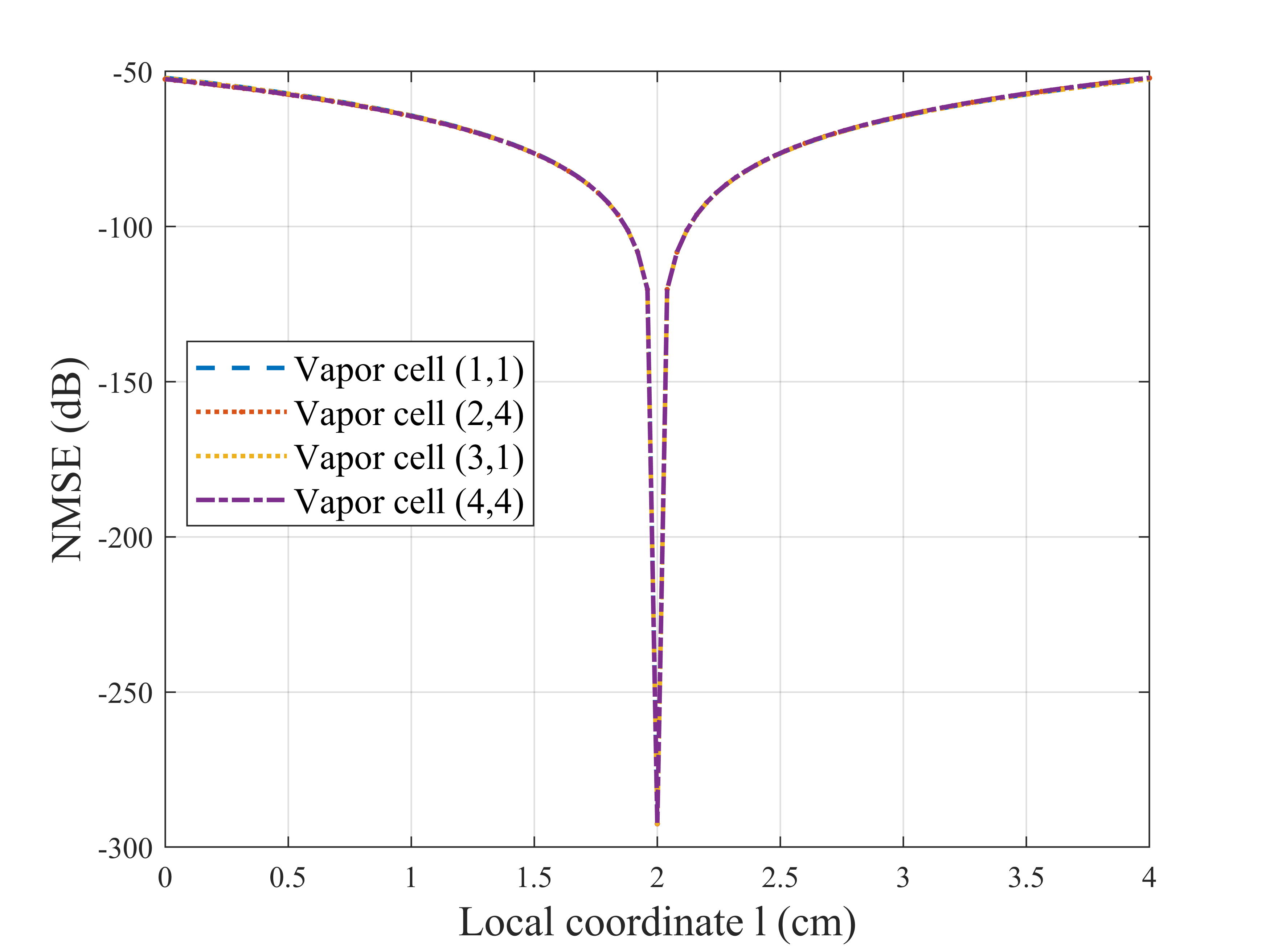}}
	\caption{Normalized mean square error between exact LO and approximated LO.}
    \label{ELOandPhase}
\end{figure*}

First, we investigate the approximation error between our derived results and the 
simulation results. Fig.~\ref{ELOandPhase}(a) and (b) depict the normalized mean square errors (NMSEs) of the amplitude in~\eqref{eq:lo_amp_center_approx} and the  phase in~\eqref{eq:lo_phase_taylor_closed}, respectively, as a function of the local coordinate \(l\). As observed, the approximation error grows with the distance from the 
center point. Moreover, the proposed center-based method accurately captures the amplitude and phase of the synthesized LO within each vapor cell.

\begin{figure}[t]
    \centering
    \includegraphics[width=1\linewidth]{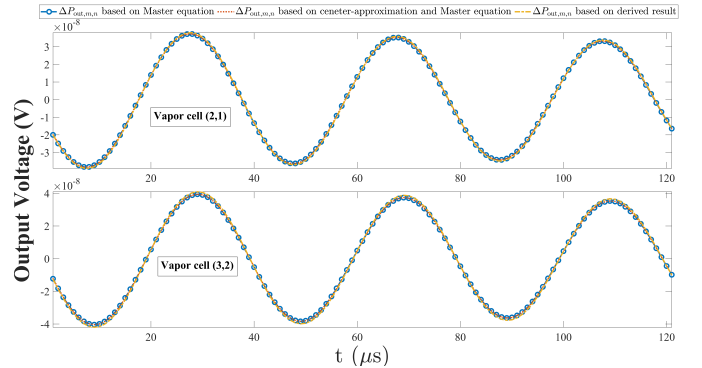}
    \caption{The output voltage based on the exact results, approximated results, and analytical derivation.}
    \label{outputvol}
\end{figure}

Then, Fig. \ref{outputvol} depicts the AC output voltage obtained using three approaches, namely the exact result from the master equation, the result from  the master equation under the center approximation, and the result from the 
closed-form analytical expression derived in this paper. As shown, the closed-form results match the master-equation results, confirming the accuracy of our derived RF-to-optical transduction.
\subsection{Cell-Level Channel Shaping}
\begin{figure*}[t]
	\centering
	\subfigure[Original beam pattern.]{
		\includegraphics[width=3.25in]{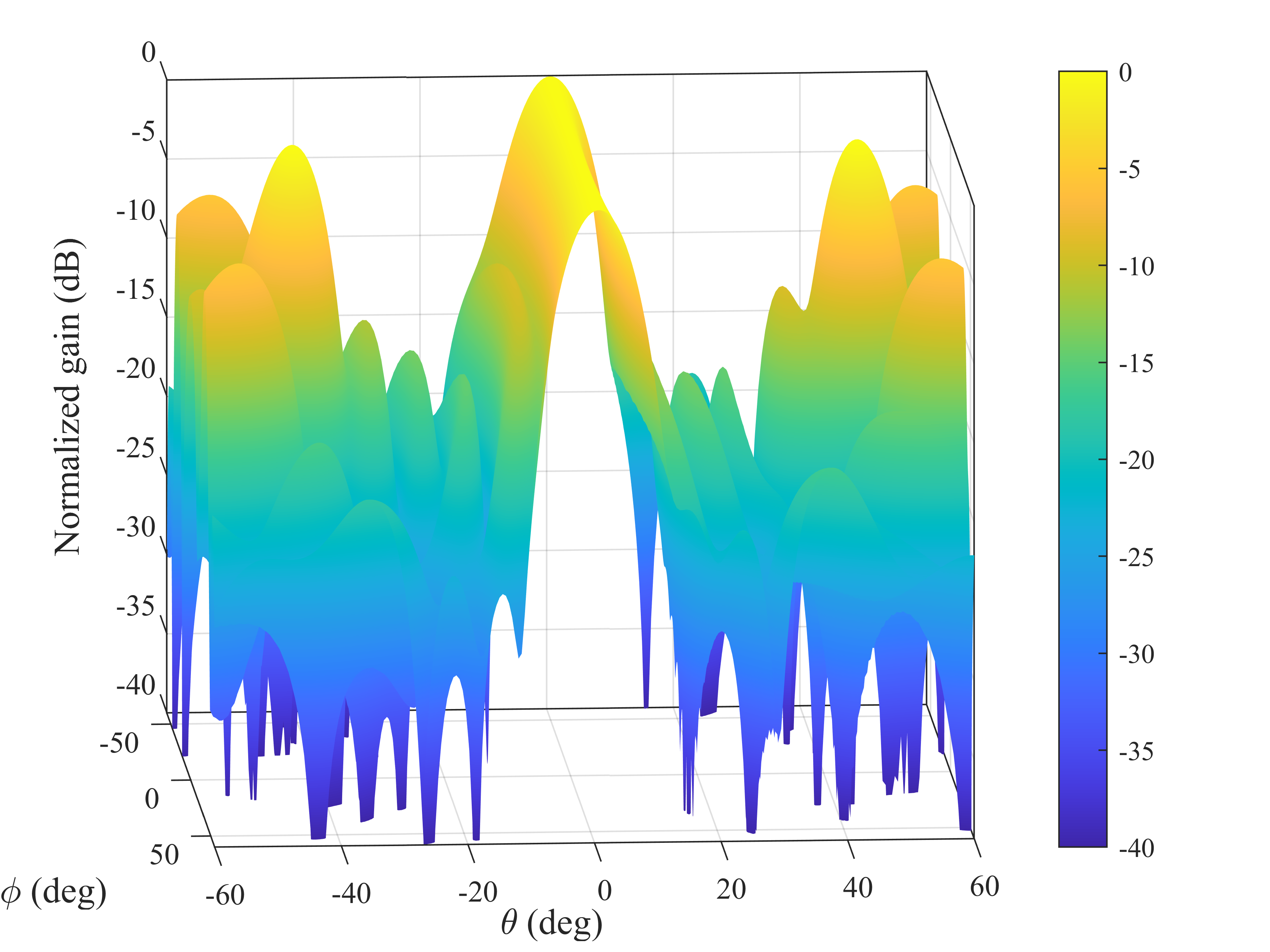}}\hspace{10mm}
	\subfigure[Beam pattern after vapor cell.]{
		\includegraphics[width=3.25in]{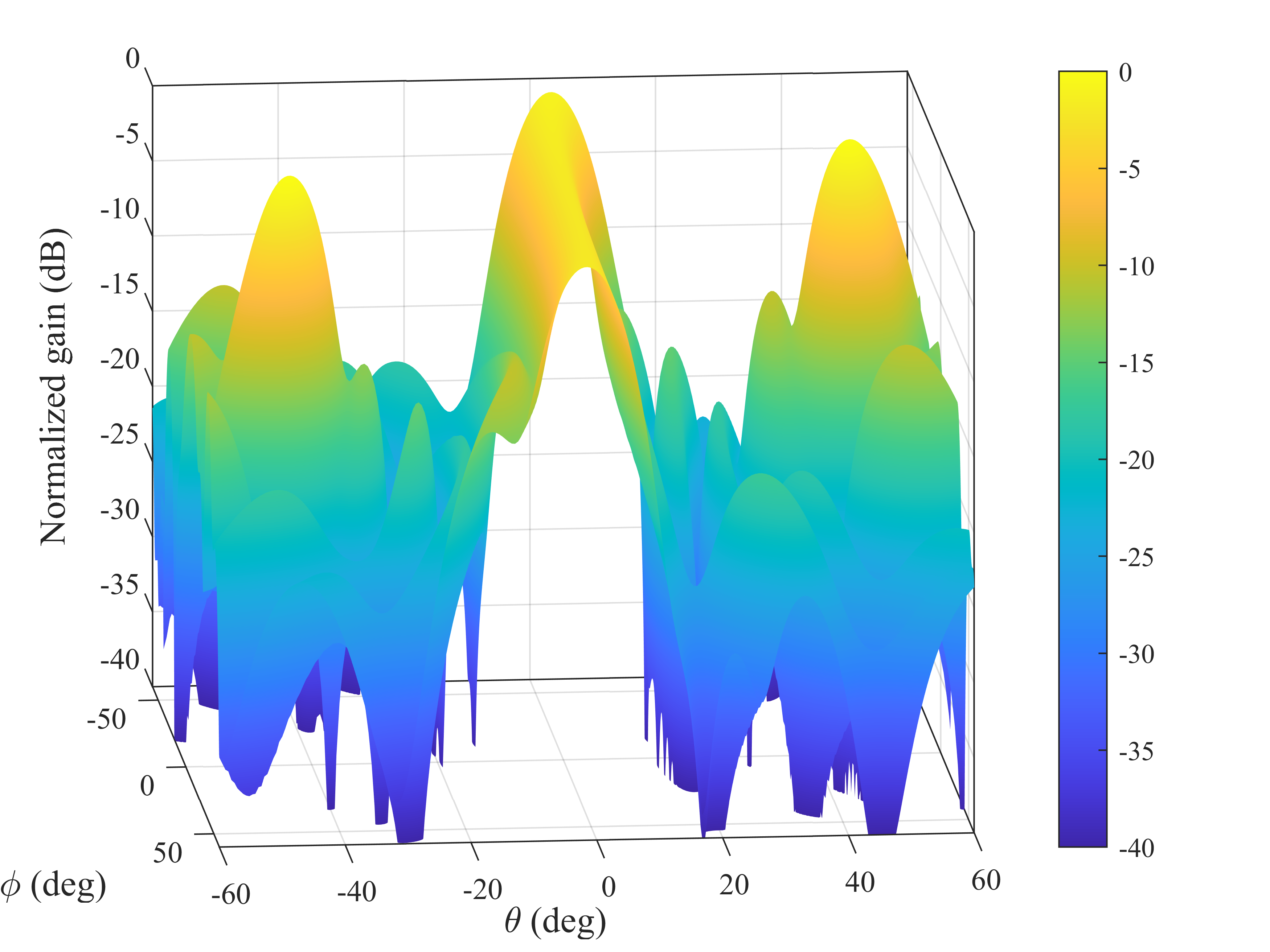}}
	\caption{3D effective beam pattern comparison.}
    \label{BP}
     \vspace{-0.2cm}
\end{figure*}

Based on the derived expression, we compare the beam pattern of the original channel with that of the effective channel in Fig.~\ref{BP}. As expected, the effective beam pattern after the vapor cell differs from that of the original channel, demonstrating the vapor cell's capability to reshape the channel and confirming our Remark \ref{rem:lo_aperture_shaping_corr}.

\subsection{Channel Correlation Relying on Vapor Cell}
\begin{figure}[t]
    \centering
    \includegraphics[width=1\linewidth]{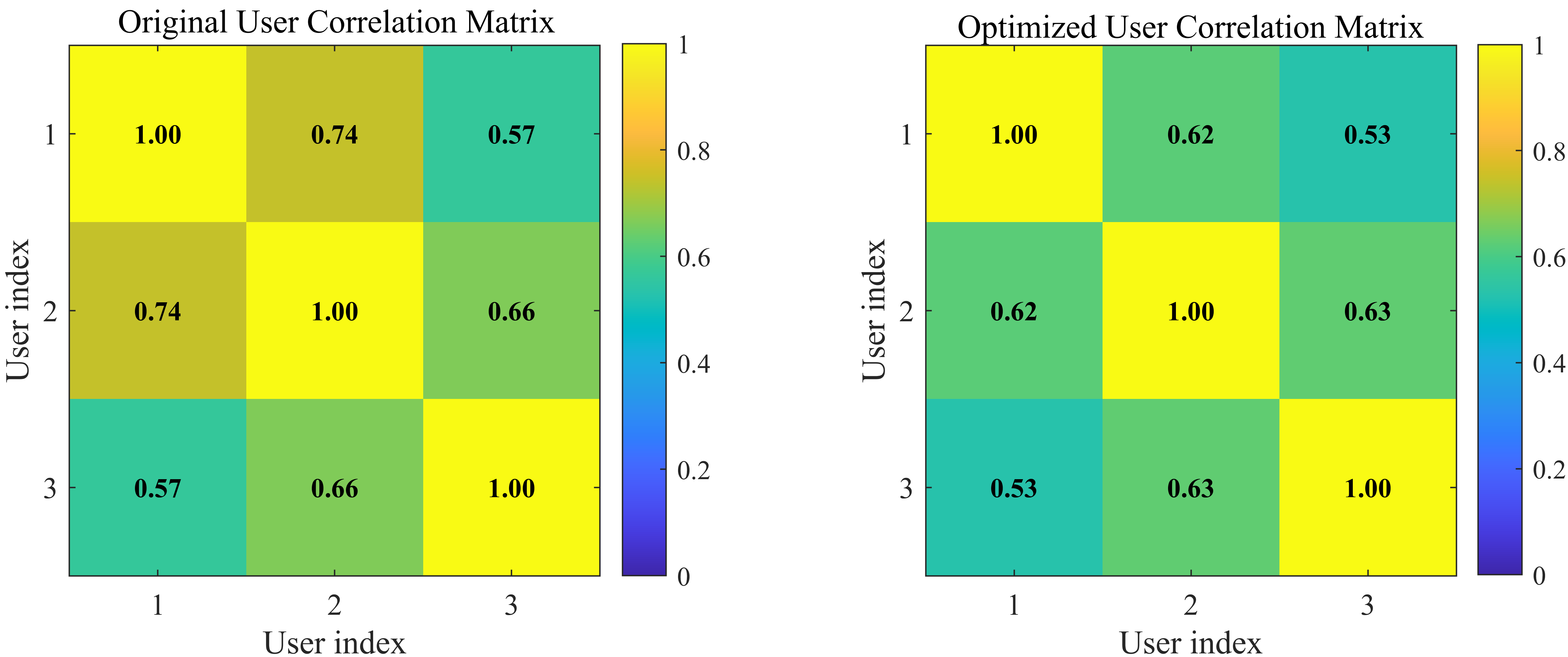}
    \caption{Correlated matrix of users' channels.}
    \label{USerCM}
     \vspace{-0.2cm}
\end{figure}

In this subsection, we investigate the spatial separation capability enabled by LO design. Fig.~\ref{USerCM} compares the correlation matrices of the original RF channel and the optimized LO design. As observed, the inter-user correlation is reduced, which benefits the system performance. However, the reduction remains limited, thereby validating Remark~\ref{LimitedCC}.

\subsection{Convergence of Proposed Algorithm}
\begin{figure}[t]
    \centering
    \includegraphics[width=1\linewidth]{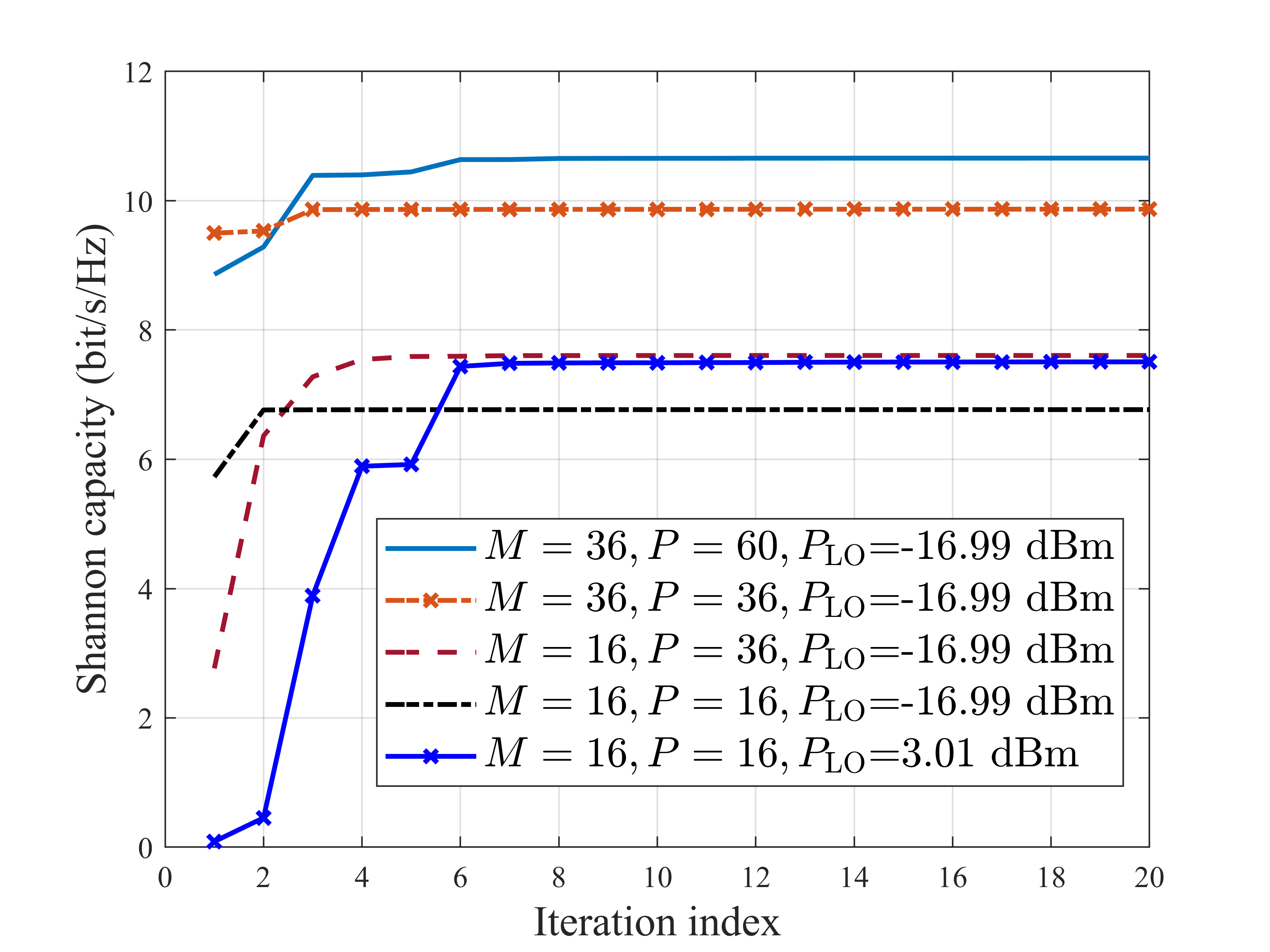}
    \caption{Convergence of our proposed algorithm.}
    \label{conver}
     \vspace{-0.2cm}
\end{figure}

To verify the convergence of the proposed algorithm, Fig.~\ref{conver} plots the Shannon 
capacity of the effective channel versus the iteration index. As can be seen, the 
algorithm converges to a locally optimal solution within 10 iterations, confirming its 
fast convergence.

\subsection{Performance Comparison}
\begin{figure}
    \centering
    \includegraphics[width=1\linewidth]{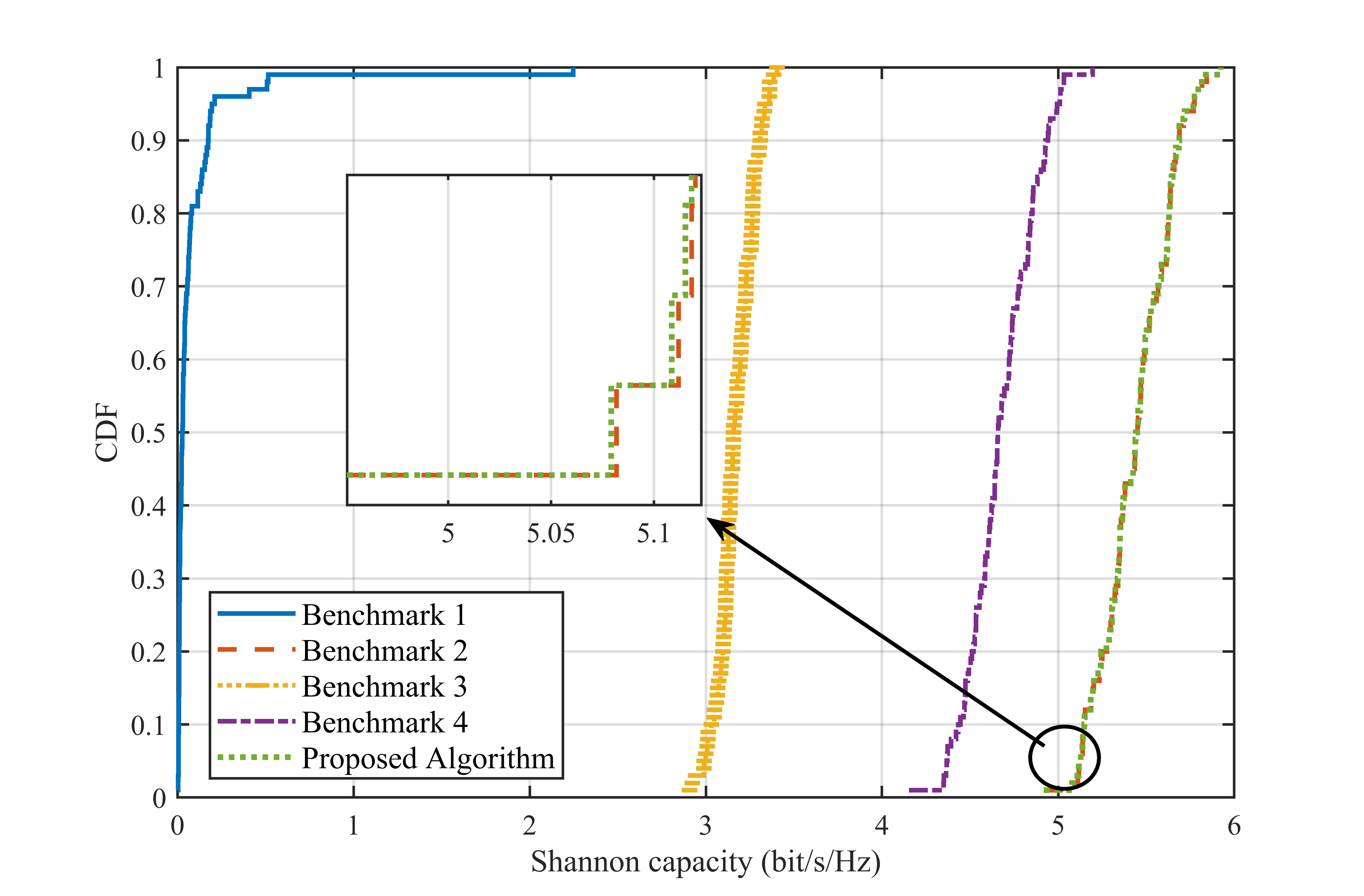}
    \caption{Performance of various algorithms with \(R_{{\rm R},k}=10\), \(\forall k\).}
    \label{Percomcdf}
    \vspace{-0.2cm}
\end{figure}

To demonstrate the effectiveness of our proposed method, the following benchmarks are presented.
\begin{enumerate}
    \item \textbf{Benchmark 1}: The amplifier coefficients and phase shifts are randomly chosen.
    \item \textbf{Benchmark 2}: We optimize the amplifier and phase shift of the LO using the genetic algorithm (GA).
    \item \textbf{Benchmark 3}: The Shannon capacity of the conventional RF receiver is adopted.
    \item \textbf{Benchmark 4}: We deploy a far-field LO and optimize power and phase shift by using our proposed method.
\end{enumerate}
Based on 100 randomly generated channel realizations, Fig.~\ref{Percomcdf} plots the cumulative distribution function (CDF) of the Shannon capacity. It can be observed that the proposed algorithm achieves performance close to that of the GA-based scheme while requiring substantially lower computational complexity. In addition, an inappropriate LO design may lead to a severe performance degradation, even worse than a conventional RF antenna. Compared with the far-field LO architecture, the near-field LO one can realize cell-level channel shaping, which confirms Corollary \ref{farfiled}.

\subsection{Effects of Vapor-Cell Length}
\begin{figure}
    \centering
    \includegraphics[width=1\linewidth]{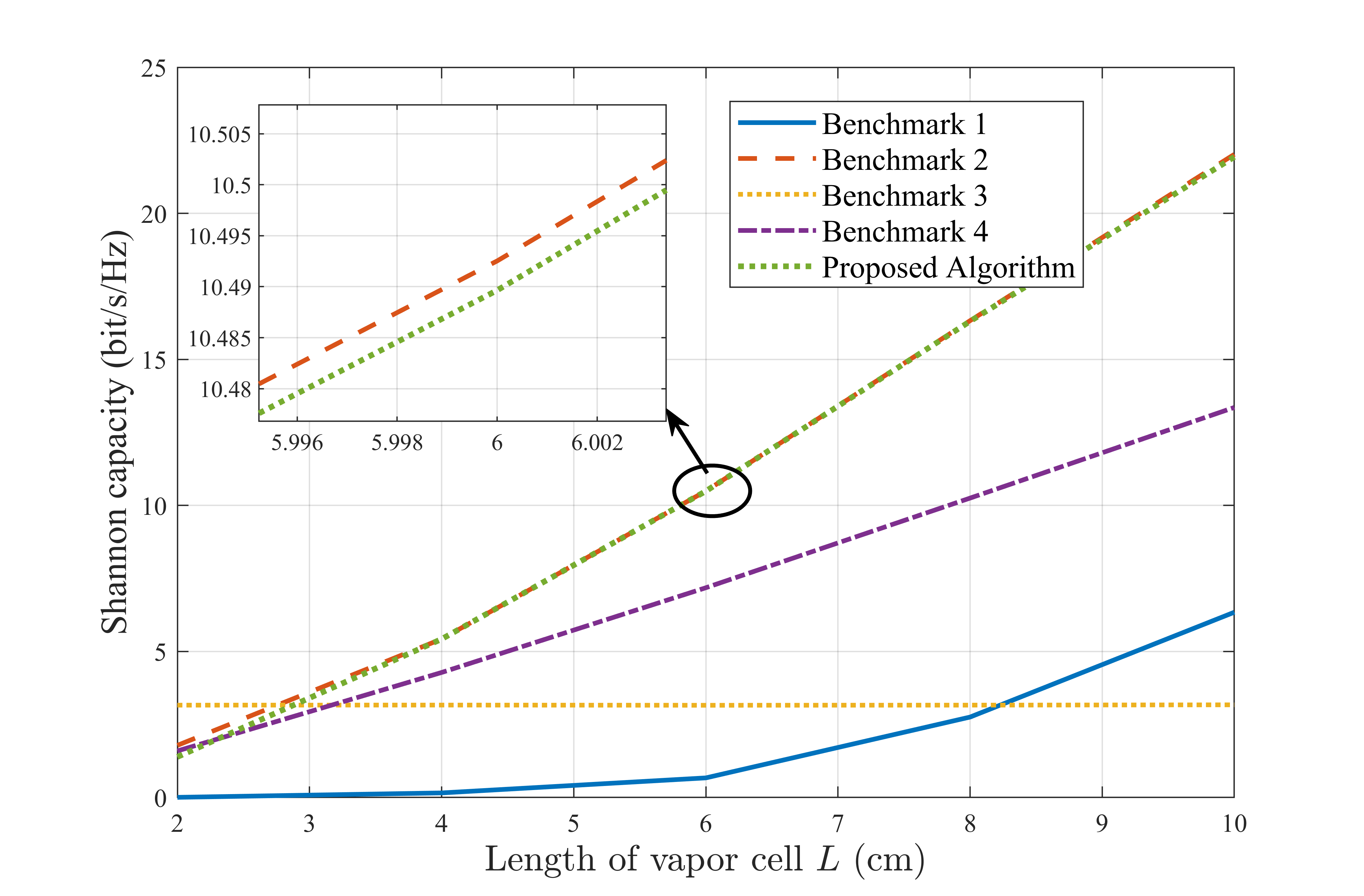}
    \caption{Performance under various Vapor cell's length \(L\).}
    \label{Perwithlength}
     \vspace{-0.2cm}
\end{figure}

In Fig.~\ref{Perwithlength}, we investigate the impact of the vapor-cell length on the Shannon capacity by averaging over 100 channel realizations. As the vapor-cell length \(L\) increases, the system capacity improves noticeably. First, a longer vapor cell provides a larger atom–light interaction region, which strengthens the accumulated RF-to-optical conversion and enhances the effective receive gain. Second, increasing  \(L\)  improves the phase-matching selectivity within each cell, making the LO-shaped response more directional and better able to discriminate the desired signal components. Finally, a larger \(L\) increases the effective aperture along the cell axis, further improving spatial separability and leading to a more favorable singular-value distribution of the effective channel.

\subsection{Effects of UPA Configuration}

\begin{figure}
    \centering
    \includegraphics[width=1\linewidth]{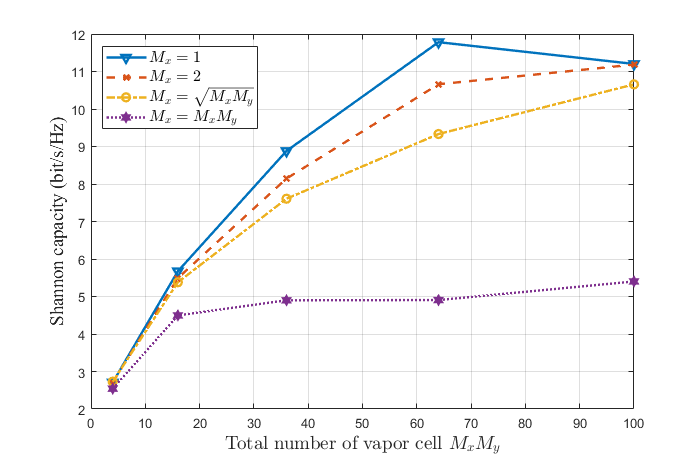}
    \caption{Performance with various UPA configurations.}
    \label{Perconfi}
     \vspace{-0.2cm}
\end{figure}

Fig.~\ref{Perconfi} reveals the strong impact of the vapor-cell layout on the Shannon capacity. Arranging the cells along the \(y\)-axis, namely \(M_x=1\), yields the highest capacity, since the half-wavelength spacing \(d_y=\lambda_c/2\) provides regular spatial sampling and stable coherent combining. In contrast, arranging all cells along the \(x\)-axis yields the worst performance. This is because the \(x\)-direction coincides with the vapor-cell longitudinal axis, and thus the 
effective cell pitch becomes \(D_x=L+d_x>\lambda_c/2\), introducing spatial aliasing and a nonuniform LO-induced response. Consequently, increasing the number of cells 
along the \(x\)-axis does not yield an efficient aperture gain.

For the \(y\)-axis arrangement, the capacity first increases with the number of cells owing to the enlarged effective aperture and improved spatial separability but then 
saturates or slightly decreases when the array becomes very long. This does not violate the array-gain intuition, but it reflects the finite LO coverage and shaping resources. As the array lengthens, the edge cells suffer from increasing LO 
near-field nonuniformity in amplitude, center phase, and local phase slope, which weakens their coherent contribution and ultimately limits the achievable gain.

\section{Conclusion}
In this paper, we proposed a self-superheterodyne Rydberg uniform array receiver for satellite uplink communications that exploits the naturally induced Doppler shift as the intermediate-frequency source. A near-field LO synthesis model was developed to characterize the spatially varying LO electric field across the Rydberg vapor-cell array, and a closed-form RF-to-optical conversion model was derived under the vapor-cell-center approximation. The resulting framework established a more accurate relationship between the incident satellite signal and the LO-induced cell-level response. Furthermore, the equivalent channel model revealed that the programmable LO can act as an analog-domain channel-shaping mechanism. Based on this insight, we proposed an efficient gradient-ascent-based LO design to maximize the Shannon capacity. Simulation results demonstrated that vapor-cell transduction can reshape the effective channel, improve receive-beam alignment, and moderately suppress inter-user interference. More importantly, the proposed LO design can achieve a significant capacity gain over benchmark schemes. These findings confirmed that a programmable LO provides a new analog-domain degree of freedom for Rydberg satellite receivers and offers a promising architecture for future satellite communication systems.

\bibliographystyle{IEEEtran}
\bibliography{myref}

\end{document}